\documentclass[a4paper,12pt,preprint,amsmath,showpacs,nofootinbib,groupedaddress,tightenlines]{revtex4-1}
\usepackage{color}
\usepackage{slashed}
\usepackage{graphicx}   
\usepackage{slashed}
\usepackage{color}
\usepackage[lofdepth,lotdepth]{subfig}
\usepackage{psfrag}

\newcommand{\als}{\alpha_s}
\newcommand{\ep}{\epsilon}

\newcommand{\nn}{\nonumber}

\begin{document}

\preprint{SLAC-PUB-15732}

\title{Single soft gluon emission at two loops}

\author{Ye Li}
\email{yli@slac.stanford.edu}
\author{Hua Xing Zhu}
\email{hxzhu@slac.stanford.edu}
\affiliation{SLAC National Accelerator Laboratory, Stanford University, Stanford, CA 94309, USA}

\begin{abstract}
We study the single soft-gluon current at two loops with two energetic
partons in massless
perturbative QCD, which describes, for example,  the soft limit of the two-loop amplitude for
$gg\to Hg$. The results are presented as Laurent expansions in
$\ep$ in $D=4-2\ep$ spacetime dimension. We calculate the expansion 
to  order $\ep^2$ analytically, which
is a necessary ingredient for Higgs
production at hadron colliders at next-to-next-to-next-to-leading order
in the soft-virtual approximation. We also give two-loop results of the
single soft-gluon current in ${\cal N}=4$ Super-Yang-Mills theory, and
find that it has
uniform transcendentality. By iteration relation of splitting
amplitudes, our calculations can determine the
three-loop single soft-gluon current to order $\ep^0$ in ${\cal N}=4$ Super-Yang-Mills
theory in the limit of large $N_c$.
\end{abstract}

\maketitle

\section{I\lowercase{ntroduction}}
\label{sec:intro}

Amplitudes in gauge theory develop infrared divergences when one or
multiple external partons become soft/collinear. Fortunately, in the
soft/collinear limit, there exist universal factorization properties
for such amplitudes, which are the foundation of higher-order
perturbative-QCD computations. Extensive discussion on the
factorization of gauge-theory amplitudes in the infrared region can be
found, for example, in
Refs.~\cite{Altarelli:1977zs,Bassetto:1984ik,Berends:1988zn,Dokshitzer:1991wu,Ellis:1991qj,Campbell:1997hg,Catani:1999ss,Bern:1995ix,Kosower:1999xi,Catani:1998nv,DelDuca:1999ha,Bern:1998sc,Catani:1998bh,Bern:1999ry,Catani:2000pi,Kosower:2003cz,Catani:2003vu,Bern:2004cz,Badger:2004uk,GehrmannDeRidder:2007jk,Bierenbaum:2011gg,Catani:2011st,Currie:2013vh,Feige:2013zla}. 

In QCD, the radiation of an arbitrary number of soft gluons off a tree-level
amplitude can be obtained using the well-known Berends-Giele recursion
relation~\cite{Berends:1988zn}. Due to the long-range properties of
soft gluon radiation, amplitudes in the soft limit have non-local
color correlations. Compact expressions for tree-level amplitudes with
two soft partons have been obtained in the color space formalism
in Ref.~\cite{Catani:1999ss}. Emission of a single soft gluon
from a generic one-loop amplitude have also been studied by several
groups~\cite{Bern:1995ix,Bern:1998sc,Bern:1999ry,Catani:2000pi}. These
results have been proven to be important in the program of next-to-next-to-leading order~(NNLO) QCD computations for jet physics, see for example
Refs.~\cite{GehrmannDeRidder:2005cm,Somogyi:2005xz,Czakon:2010td,Baernreuther:2012ws,Czakon:2013goa,Ridder:2013mf,Boughezal:2013uia}.

While the NNLO revolution is under way, there is strong motivation for
going one order in $\als$ further. This is driven by both experimental
and theoretical demands. On the experimental side, the discovery of
Higgs boson marks one of the most important progress in particle physics in
the last few decades~\cite{Aad:2012tfa,Chatrchyan:2012ufa}. It's
certainly important to give the most precise theoretical prediction for
its production cross section. On the theory side, uncertainties
estimated by scale variation for Higgs production is around $\pm 10\%$
at
NNLO~\cite{Harlander:2002wh,Anastasiou:2002yz,Ravindran:2003um,Catani:2007vq},
and improved to $\pm 7\%$ by including soft-gluon resummation up to
next-to-next-to-leading logarithmic accuracy~\cite{Dittmaier:2011ti}. Further decreasing the scale uncertainties to percent level requires the
computation of next-to-next-to-next-to-leading order~(NNNLO) QCD
corrections.

In this paper we consider single soft-gluon radiation at two loops,
which plays an important role in NNNLO
QCD corrections, similar to the one-loop soft-gluon current does in
NNLO computations, see for example, Refs.~\cite{GehrmannDeRidder:2005cm,Somogyi:2005xz,Czakon:2010td,Boughezal:2011jf}. To
simplify the situation, we confine ourselves to the case that only two hard partons
are present. This corresponds to the
cases such as $e^+e^-\to$ dijet, deep-inelastic scattering, or
Drell-Yan/Higgs production at hadron collider. Previously, such amplitudes have been
derived~\cite{Badger:2004uk} by taking the soft limit of collinear splitting amplitudes at
two loops to order $\ep^0$, using the two-loop aamplitudes for
$\gamma^* \to q\bar{q}g$~\cite{Garland:2001tf,Garland:2002ak} and $H\to ggg$~\cite{Gehrmann:2011aa}. However,
for a NNNLO computation, one needs
the Laurent expansion in $\ep$ through the $\ep^2$
terms, which we have given for the first time in this paper.
 Our results for the two-loop soft amplitude  agree with the soft
 limit of the two-loop splitting amplitudes~\cite{Bern:2004cz,Badger:2004uk} through the $\ep^0$ terms, serving as a strong check of our
calculation.

As a by-product, we obtain the soft-gluon
current in ${\cal N}=4$ Super-Yang-Mills theory to order
$\ep^2$, which coincides with the QCD result at leading
transcendentality. We also derive the soft limit of splitting
amplitudes at three loops through order $\ep^0$ at leading order of
$N_c\to \infty$, using the results of Refs.~\cite{Bern:2005iz,Spradlin:2008uu}.

The paper is organized as follows. In Sec.~\ref{sec:1}, we review the general
result on the factorization of the single
soft-gluon current at tree level and one loop. In Sec.~\ref{sec:3} we
 calculate the soft-gluon current to two loops. We
conclude at Sec.~\ref{sec:5}. We present some details for the computation of one of the master integral in the appendix.

\section{R\lowercase{eview of the soft-gluon current}}
\label{sec:1}

In this section we review the factorization of amplitudes in the soft
limit, closely following the notation in
Ref.~\cite{Catani:2000pi}.
It's well-known~\cite{Bassetto:1984ik,Dokshitzer:1991wu} that
tree-level QCD amplitudes with two hard partons and one soft gluon can
be written as
\begin{eqnarray}
\label{eq:1}
|\mathcal{M}^{(0)} (q,p_1,p_2)|^2 &\simeq& 
4g^2_s \mu^{2\ep} C_R
S^{(0)}_{12}(q) |\mathcal{M}^{(0)}
(p_1,p_2)|^2
\end{eqnarray}
where $S^{(0)}_{12}(q)=\frac{p_1\cdot p_2}{2(q\cdot p_1)(q\cdot
  p_2)}$, and $\mathcal{M}^{(0)} (q,p_1, p_2) $ is the
tree-level amplitude for $2$ hard partons~(massless quark or gluon) and one soft gluon, and
$\mathcal{M}^{(0)}(p_1,p_2)$ is the corresponding amplitude with
the soft gluon stripped off. Dependence of the amplitudes on the extra colorless
particles in the process is left implicit. The symbol $\simeq$ means that we have neglected terms that are less
singular than $1/q^2$. $g_s$ is the strong coupling constant, $\mu$ is the
mass scale introduced by continuing the space-time dimension to
$D=4-2\ep$ dimension. $C_R$ is the quadratic Casimir invariant. $C_R=C_A$ if
parton $1$ is a gluon, $C_R=C_F$ if parton $1$ is a quark, where $C_A=N_c$ and $C_F=\frac{N^2_c-1}{2N_c}$, with $N_c$ being the number of color. Note that
the functional dependence of the eikonal function $S^{(0)}_{12}(q)$ is
uniquely determined by its invariance under the rescaling of
$p_1$ and $p_2$, which is a simple consquence of the QCD Feynman rule in the
eikonal limit. 
In our convention, all momenta are massless and
have positive-definite energies. The generalization of Eq.~(\ref{eq:1}) to
processes with any
number of hard partons can be found, for example, in
Ref.~\cite{Catani:1999ss}. 

At the one-loop level, Eq.~(\ref{eq:1}) receives quantum corrections,
which can be written as
\begin{eqnarray}
&&  \mathcal{M}^{(0)} (q,p_1, p_2) \mathcal{M}^{(1)} (q,p_1,
  p_2)^*  +\mathrm{c.c.}
\nn
\\
&\simeq&
  \left(4 (g_s\mu^\ep)^2 C_R  S^{(0)}_{12}(q)\mathcal{M}^{(0)} (p_1, p_2) \mathcal{M}^{(1)} (p_1,
  p_2)^*  + \mathrm{c.c.} \right)
\nn
\\
&&
+\left(4 (g_s\mu^\ep)^2 C_R S^{(1)}_{12}(q) |\mathcal{M}^{(0)}
(p_1,p_2)|^2 + \mathrm{c.c.}\right),
\label{eq:2}
\end{eqnarray}
where $\mathrm{c.c.}$ denotes complex conjugate. $\mathcal{M}^{(i)}$
is the $i$th order in $\als$ unrenormalized amplitudes in dimensional
regularization, where UV and IR divergences are simutaneously
regularized by the dimensional regularization parameter $\ep$.  The one-loop corrections to the eikonal function have been
calculated to
be~\cite{Bern:1995ix,Bern:1998sc,Bern:1999ry,Catani:2000pi}
\begin{eqnarray}
  \label{eq:3}
  S^{(1)}_{12}(q) = - S^{(0)}_{12} (q)\frac{\als}{4\pi}  C_A S_\ep
\frac{e^{\ep \gamma_E}\Gamma^3(1-\ep)\Gamma^2(1+\ep)}{\ep^2
    \Gamma(1-2\ep)},
\end{eqnarray}
where
\begin{eqnarray}
  S_\ep =\left( 4\pi e^{-\gamma_E}
      e^{i\sigma_{12} \pi} \mu^2 S^{(0)}_{12}(q) \right)^\ep,
\end{eqnarray}
and $\sigma_{12}=-1$ if both $p_1$ and $p_2$ are incoming, otherwise $\sigma_{12}=1$.
Note that the one-loop eikonal
function doesn't depend on $C_R$, which may be explained by the non-abelian exponentiation theorem~\cite{Gatheral:1983cz,Frenkel:1984pz},
if one replaces the polarization summation for the soft gluon by a cut propagator.
Eq.~(\ref{eq:3}), and its
generalization to processes with any number of hard partons  have  been
used, for example, in the calculation of soft-virtual approximation to Higgs
production at NNLO~\cite{Catani:2001ic,Harlander:2001is,deFlorian:2012za}, in the
calculation of the two-loop soft
function in soft-collinear-effective theory~\cite{Bauer:2000ew,Bauer:2000yr,Bauer:2001yt,Beneke:2002ph}, and in the construction of subtraction term in general NNLO corrections~\cite{GehrmannDeRidder:2005cm,Somogyi:2005xz,Czakon:2010td,Boughezal:2011jf}.

\section{C\lowercase{alculation of the two-loop soft-gluon current}}
\label{sec:3}

At two loops, the factorized soft-gluon current has the form 
\begin{eqnarray}
&&  \mathcal{M}^{(0)} (q,p_1, p_2) \mathcal{M}^{(2)} (q,p_1,
  p_2)^*  +\mathrm{c.c.}
\nn
\\
&\simeq&
  4 (g_s\mu^\ep)^2 C_R  \Big[ \left(S^{(0)}_{12}(q)\mathcal{M}^{(0)} (p_1, p_2) \mathcal{M}^{(2)} (p_1,
  p_2)^*  + \mathrm{c.c.} \right)
\nn
\\
&&+
 \left(  S^{(1)}_{12}(q)\mathcal{M}^{(0)} (p_1, p_2) \mathcal{M}^{(1)} (p_1,
  p_2)^*  + \mathrm{c.c.} \right)
\nn
\\
&&
+\left( S^{(2)}_{12}(q) |\mathcal{M}^{(0)}
(p_1,p_2)|^2 + \mathrm{c.c.}\right) \Big].
\label{eq:twofac}
\end{eqnarray}
The two-loop generalization is consistent with the soft limit of
two-loop collinear splitting
amplitudes~\cite{Bern:2004cz,Badger:2004uk}. For the latter, it has
been shown that similar factorization form holds to all orders in
$\als$~\cite{Kosower:1999xi}. The two-loop eikonal function,
$S^{(2)}_{12}(q)$, is known through the order $\ep^0$ terms by taking the
soft limit of the two-loop collinear splitting
amplitudes~\cite{Bern:2004cz,Badger:2004uk} or the two-loop squared
amplitudes for $\gamma^* \to q
\bar{q} g$ and $H\to ggg$~\cite{Garland:2001tf,Garland:2002ak,Gehrmann:2011aa}. However, for computation
accurate to NNNLO, one also needs the order $\ep$ and order $\ep^2$
terms. In this section, we
calculate the Laurent expansions of $S^{(2)}_{12}(q)$ in $\ep$ through
order $\ep^2$, using a method different from
Refs.~\cite{Badger:2004uk}.

In Ref.~\cite{Catani:2000pi}, the one-loop soft-gluon current is
derived by taking the eikonal approximation of the integrand of the
amplitudes before the loop integrals are carried out. This has the
advantage that the one-loop eikonal function can be directly obtained without
the subtraction of the product of the tree-level eikonal function and
the one-loop squared amplitude, that is, the second line of
Eq.~(\ref{eq:2}). The same procedure can be used in the calculation of
the two-loop eikonal function.

Specifically, we generate the integrand corresponding to the
interference of tree-level and two-loop amplitudes, the first line of
Eq.~(\ref{eq:twofac}). For this purpose, we consider the process
$\gamma^* \to q(p_1)\bar{q}(p_2)g(q)$, keeping in mind that the eikonal function is
independent of the colorless particles in the process. Summation of
polarization for the external gluon is done in Feynman gauge. We then take
the eikonal approximation of the integrand, assumming that the
energy of the internal and external gluons are parametrically smaller then
$p^0_1$ and $p^0_2$. The integrand after the eikonal
approximation is taken can also be generated by treating $q(p_1)$ and
$q(p_2)$ as two out-going Wilson lines, whose directions are given by
$p^\mu_1/p^0_1$ and $p^\mu_2/p^0_2$. We have checked that this indeed
gives the same integrand~\footnote{We use QGRAF~\cite{Nogueira:1991ex} extensively in
generating various diagrams.}. We note that after the eikonal approximation is
taken on the right hand side of Eq.~(\ref{eq:twofac}), the second and the third lines of
it vanish. The reason is that $\mathcal{M}^{(1)}(p_1,p_2)$ and
$\mathcal{M}^{(2)}(p_1,p_2)$ become scaleless integrals and vanish
identically in dimensional regularization. Therefore, 
The two-loop eikonal function can
be obtained by evaluating the resulting integrand, without the need of
subtraction.





\subsection{Warm up: one-loop soft-gluon current}
\label{sec:3a}

\begin{figure}[ht]
  \centering
  \includegraphics[width=0.2\linewidth]{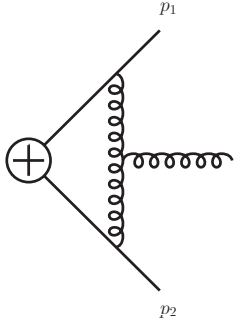}
  \caption{Non-vanishing diagram for soft gluon emission at
    one-loop. Solid line are quark/anti-quark lines in the high energy
  limit.}
\label{fig:oneloop}
\end{figure}

\begin{figure}[ht]
  \centering
  \includegraphics[width=0.2\linewidth]{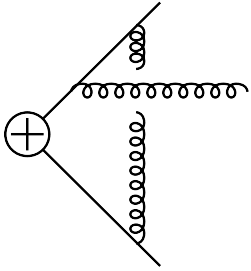}
  \caption{Diagram which vanishes in dimensional regularization.}
\label{fig:oneloopzero}
\end{figure}

For the convenience of reader,
we reproduce the one-loop results in this section. At one loop, there
is only one non-zero diagram~(from now on eikonal approximation is
assumed for the integrand), which is depicted
in Fig.~\ref{fig:oneloop}. All the remaining diagrams are zero in
dimensional regularization, because their loop integrals are
scaleless. One example of such a vanishing diagram is depicted in
Fig.~\ref{fig:oneloopzero}. We note that the
external soft-gluon momentum only enters the loop integral through $q\cdot
p_1$. However, the invariance of the integral under the rescaling of $p_1$ and $p_2$
demands that a factor of $\left(\frac{\mu^2 (p_1\cdot p_2)}{(q\cdot p_1)
  (q\cdot p_2)}\right)^\ep$ must be generated per loop. This is
impossible for this diagram, leading to the conclusion that it must vanish. 

\begin{figure}[ht]
  \centering
  \includegraphics[width=0.4\linewidth]{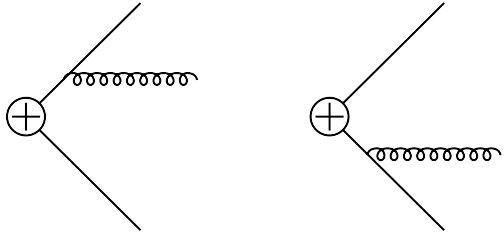}
  \caption{Tree-level diagrams for single soft gluon emission.}
\label{fig:tree}
\end{figure}
We calculate the interference between the one-loop non-zero diagram,
Fig.~\ref{fig:oneloop}, and the tree-level diagrams in
Fig.~\ref{fig:tree}. The one-loop eikonal function can then be extracted from the
one-loop integral in the interference term, which, after some
simplification, reads
\begin{eqnarray}
\label{eq:4}
S^{(1)}_{12}(q) =  i 4 g^2_s C_A (p_1\cdot p_2) \mu^{2\ep}
\int\!\frac{d^D k}{(2\pi)^D} \frac{1}{[2k\cdot p_1] [2(q-k)\cdot p_2]
  [k^2] [(k-q)^2]},
\end{eqnarray}
where the Feynman prescription $i0^+$ is implicitly understood for all propagators in square
brackets, for example, $[k^2]\equiv k^2 + i0^+$. Carrying out the loop integral, we reproduce the one-loop
eikonal function in Eq.~(\ref{eq:3}).

\subsection{Two-loop soft-gluon current}
\label{sec:3b}

\begin{figure}[]
  \centering
  \includegraphics[width=0.8\linewidth]{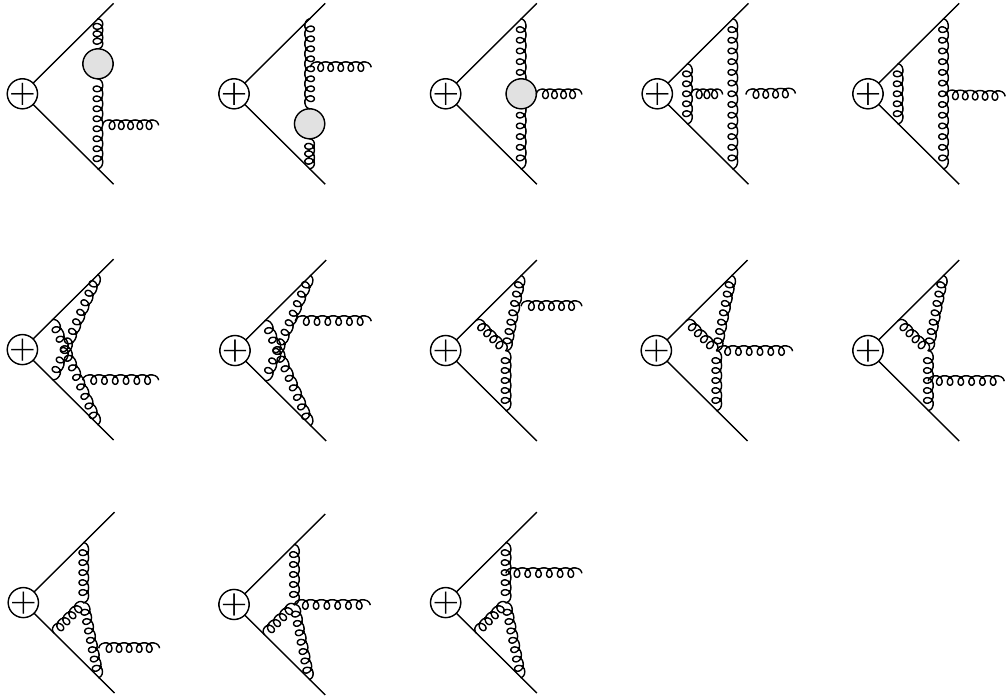}
  \caption{Two-loop non-vanishing diagrams for single soft gluon emission.}
\label{fig:twoloop}
\end{figure}

\begin{figure}[]
  \centering
\subfloat[][]{
  \includegraphics[width=0.15\textwidth]{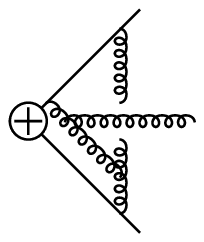}
\label{fig:0a}
}
\qquad
\subfloat[][]{
  \includegraphics[width=0.15\textwidth]{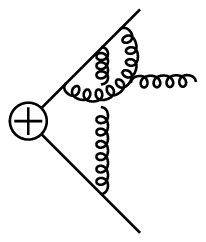}
\label{fig:0b}
}
  \caption{Examples of diagrams which vanish identically.}
\end{figure}

As explained above,
the two-loop eikonal function $S^{(2)}_{12}(q)$ can be extracted from
the calculation of the non-vanishing diagrams at two-loop level, as depicted in
Fig.~\ref{fig:twoloop}. The grey blobs represent all possible
two-point and three point insertions,
where no eikonal approximation is made. We include $N_f$ flavour of massless fermions and $N_s$ flavour of massless scalar in the blob, besides the gluon. In QCD, $N_f=5$, $N_s=0$.
Before describing the calculation of these diagrams, we
comment on the diagrams that vanish identically. There are two
classes of vanishing diagrams. The first class vanishes due to color
or Lorentz algebra. An example of it is depicted in
Fig.~\ref{fig:0a}. The second class vanishes because the
corresponding loop integral is scaleless, as in
Fig.~\ref{fig:0b}. Because of the vanishing of these two classes of
diagrams, the actual number of diagrams that need to be evaluated is
significantly reduced.

\begin{figure}[]
  \centering
\subfloat[sub-caption][$I_1$]{
  \includegraphics[width=0.25\textwidth]{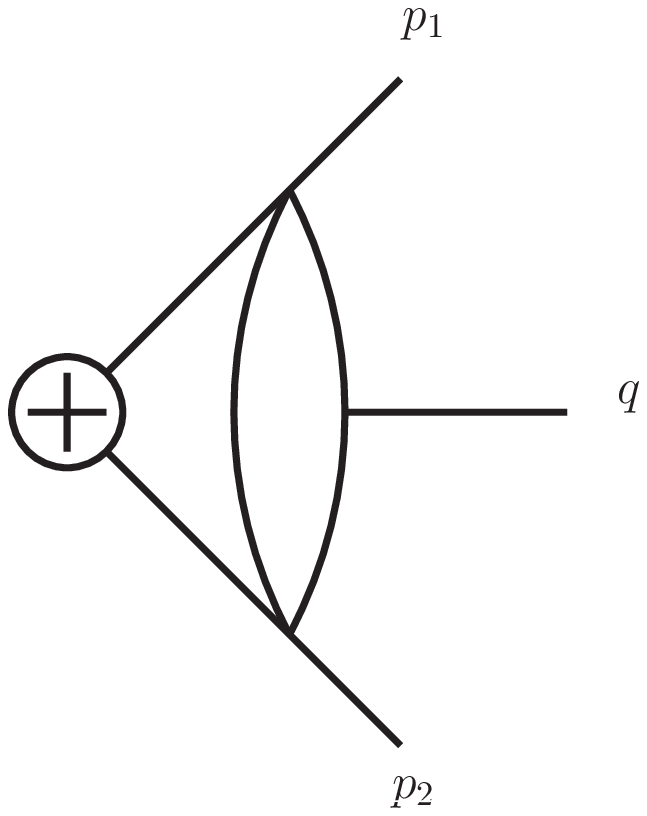}
}
\qquad
\subfloat[][$I_2$]{
  \includegraphics[width=0.25\textwidth]{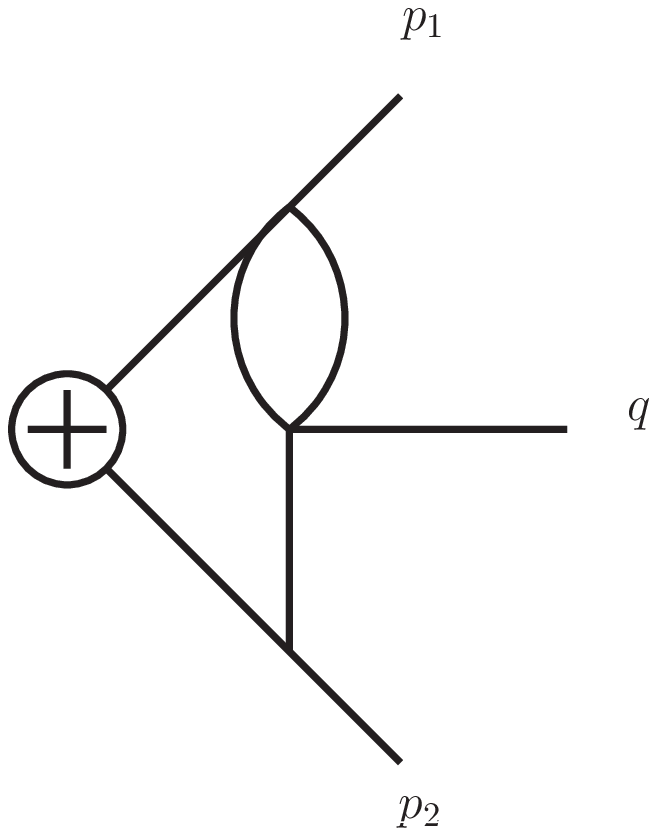}
}
\qquad
\subfloat[][$I_3$]{
  \includegraphics[width=0.25\textwidth]{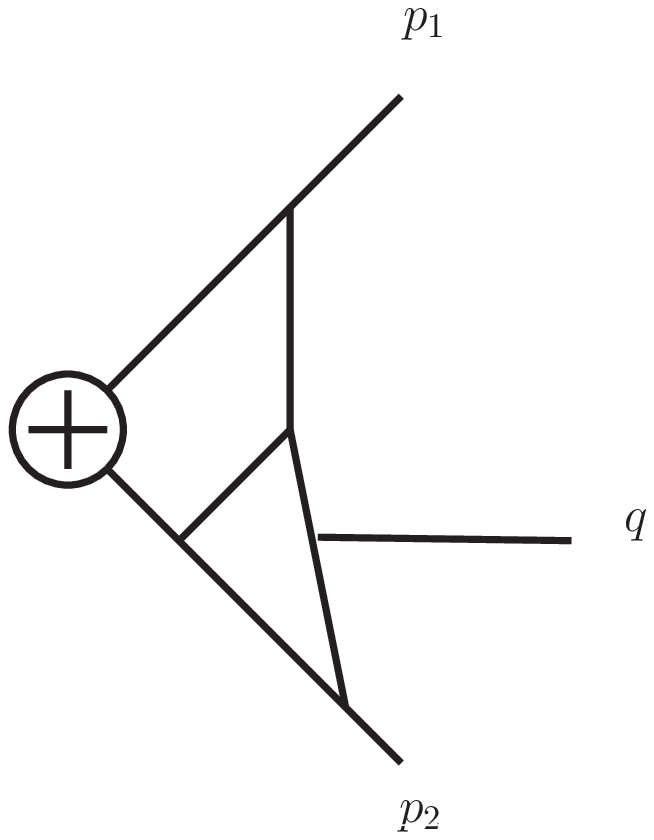}
}
\caption{Master integrals encountered in the computation. Eikonal
    approximations are taken on the directions $p_1$ and $p_2$.}
\label{fig:master}
\end{figure}

We now come to the actual evaluation of the non-zero diagrams in
Fig.~\ref{fig:twoloop}. We calculate the interference terms between the
tree-level diagrams in Fig.~\ref{fig:tree} and the two-loop diagrams
in Fig.~\ref{fig:twoloop}. After the evaluation of color factor and
kinematical factor, the resulting loop integrals are reduced to three
master integrals in Fig.~\ref{fig:master}. To that end, we use the techniques of
Integration-By-Parts~(IBP)~\cite{Chetyrkin:1981qh,Tkachov:1981wb},
implemented in the MATHEMATICA package FIRE~\cite{Smirnov:2008iw}
using the Laporta algorithm~\cite{Laporta:2001dd}. The reduction to master
integrals has also been cross checked using a different MATHEMATICA
package LiteRed~\cite{Lee:2012cn}. The results after the IBP reduction
procedure can be written as
\begin{eqnarray}
  S^{(2)}_{12}(q) &=& g^4_s \frac{p_1\cdot
      p_2}{(q\cdot p_1)(q\cdot p_2)} \times \Bigg\{
  C_A N_f \Bigg[ \frac{2(-7+2D)(12-6D+D^2) }{(-6+D)(-3+D)(-2+D)(-1+D)} I_1 
\nn
\\
&&
- \frac{6(-4+D)^2 }{ (-6+D)(-2+D)(-1+D)  } I_2 \Bigg]
+  C_A N_s \Bigg[ -
\frac{  (-7+2D) (-4-4 D + D^2)  }{2 (-6+D) (-2+D)
  (-1+D)  } I_1 
\nn
\\
&&
+ \frac{3 (-4+D)^2   }{(-6+D) (-2+D) (-1+D)  } I_2 \Bigg]
+   C^2_A \Bigg[
+\frac{8}{3}  I_3
\nn
\\
&&
- \frac{   (2 (-156 + D (72 + D (11 + (-9 + D) D))) - 3 (-4 + D)^3
  D_s)  }{ (-6+D)(-4+D)(-2+D)(-1+D) } I_2
\nn
\\
&&
+ \Bigg(
\frac{ (-7 + 2 D) (504 - 1308 D + 874 D^2 - 213 D^3 + 
    17 D^4) }{3 (-6 + D) (-4 + D) (-3 + D) (-2 + D) (-1 + D)}
\nn
\\
&&
- \frac{(-7 + 2 D) (-4 - 4 D + D^2) D_s}{2(-6 + D) (-2 + D) (-1 + D)}
\Bigg)I_1
\Bigg]\Bigg\},
\label{eq:reduction}
\end{eqnarray}
The
parameter $D_s$ selects the particular variant of dimensional
regularization. For $D_s=4-2\ep$ the scheme is the conventional
dimensional regularization scheme, while for $D_s=4$ it is the
four-dimensional helicity scheme~(FDH)~\cite{Bern:1991aq,Bern:2002zk}.

There are three master integrals encountered in this computation.
They are defined as
\begin{eqnarray}
  I_1 &=& \mu^{4\ep}\int\! \frac{d^D k_1 d^D k_2}{(2\pi)^{2D}} 
\frac{p_1\cdot p_2}{[2k_1\cdot p_1][2(q - k_1)\cdot
  p_2][k^2_2][(k_2-q)^2][(k_1-k_2)^2]},
\nn
\\
  I_2 &=& \mu^{4\ep}\int\! \frac{d^D k_1 d^D k_2}{(2\pi)^{2D}} 
\frac{p_1\cdot p_2}{[2k_1 \cdot p_1] [2(q-k_1)\cdot p_2] [k^2_1] [k^2_2] [(k_1+k_2-q)^2]},
\nn
\\
  I_3 &=& \mu^{4\ep}\int\! \frac{d^D k_1 d^D k_2}{(2\pi)^{2D}} 
\frac{(q\cdot p_1)(q\cdot p_2)^2}{[2k_1\cdot p_1][2(q-k_1)\cdot p_2] [2(k_2+q)\cdot p_2]
  [k^2_2] [(k_1+k_2)^2] [(k_2+q)^2]},
\end{eqnarray}
where $i0^+$ dependences in the propagators are understood. The first
two masters are calculated to all orders in $\ep$. For the last master
integral, we give the Laurent expansion of it to order $\ep^2$, which is the order relevant for NNNLO
computation. The details of the computation of the last integral are
presented in the appendix. Here we only list the results for the three
master integrals:
\begin{eqnarray}
  I_1 &=& -\frac{1}{(16\pi^2)^2} S^2_\ep    \frac{
    e^{2\ep\gamma_E}\Gamma^2(1-2\ep)\Gamma^2(1-\ep)  \Gamma^2(1+2\ep) }{8\ep^3 (1-4\ep)\Gamma(1-4\ep)},
\nn
\\
I_2&=& -\frac{1}{(16\pi^2)^2} S^2_\ep  \frac{ e^{2\ep\gamma_E} 
    \Gamma(1-2\ep) \Gamma^3(1-\ep) \Gamma^2(1+2\ep) }{8\ep^3 (1-2\ep) \Gamma(1-3\ep)},
\nn
\\
I_3 &=&  -\frac{1}{(16\pi^2)^2} S^2_\ep \Bigg[ 
-\frac{1}{8\ep^4} - \frac{5\zeta_2}{16\ep^2} + \frac{25\zeta_3}{48 \ep}
- \frac{17\zeta_4}{16} + \ep\left(  \frac{67\zeta_2\zeta_3}{48} +
  \frac{319\zeta_5}{80} \right)
\nn
\\
&&
+ \ep^2 \left( \frac{101\zeta^2_3}{36} + \frac{1723\zeta_6}{256} 
\right)+ \mathcal{O}(\ep^3)\Bigg],
\end{eqnarray}
where $\zeta_s$ is the Riemann
zeta vaule,  $\zeta_s = \sum^{\infty}_{n=1} \frac{1}{n^s}$.
It's interesting to note that $I_3$ coincides with the soft limit of the corresponding
master integral in full QCD, where no eikonal approximation is taken
in the denominator. The latter was calculated in
Ref.~\cite{Gehrmann:2000zt} to order $\ep^0$. This is probably due to
the fact that the divergences in $I_3$ have only infrared
origin. While we have only presented the Laurent expansions of $I_3$
to order $\ep^2$ analytically, the higher-order terms can easily be obtained
numerically, using its two-fold Mellin-Barnes integral representation
derived in the appendix, and the {\tt MBintegrate} routine of
Czakon~\cite{Czakon:2005rk}. For example, the next three terms in the
$\ep$ expansion of $I_3$ are
given by
\begin{eqnarray}
  (82.1443689 \pm 0.0000007)\ep^3 + (198.904248 \pm 0.000002)\ep^4 +
  (726.325910 \pm 0.000007) \ep^5.
\end{eqnarray}
However, it's difficult to convert them into Riemann zeta values due
to lack of significant digits.

Substituting the master integral into Eq.~(\ref{eq:reduction}) and setting
$N_s=0$, we obtain the  two-loop eikonal function in QCD in the conventional dimensional regularization scheme~($D=D_s=4-2\ep$),
\begin{eqnarray}
  S^{(2)}_{12}(q) &=& S^{(0)}_{12} (q)\left(\frac{\als }{4\pi}
 \right)^2   S^2_\ep 
\Bigg\{ C_A N_f \Bigg[ \frac{1}{6\ep^3} + \frac{5}{18\ep^2} +
\frac{19}{54\ep} + \frac{\zeta_2}{6\ep}+  \frac{65}{162}
+\frac{5\zeta_2}{18} - \frac{31\zeta_3}{9}
\nn
\\
&& + \ep \left( -\frac{35
    \zeta_2 }{54}-\frac{155
   \zeta_3}{27}-\frac{185
   \zeta_4}{24}+\frac{211}{486}  \right)
\nn
\\
&&
+\ep^2 \left( 
-\frac{31}{9} \zeta _3 \zeta _2-\frac{367
   \zeta _2}{162}-\frac{994 \zeta
   _3}{81}-\frac{925 \zeta _4}{72}-\frac{511
   \zeta _5}{15}+\frac{665}{1458}
\right)\Bigg]
\nn
\\
&&
+ C^2_A \Bigg[ \frac{1}{2\ep^4} - \frac{11}{12\ep^3} + \frac{
  -\frac{67}{36} + \zeta_2 }{\ep^2} + \frac{ -\frac{193}{54} -
  \frac{11\zeta_2}{12} - \frac{11\zeta_3}{6} }{\ep} - \frac{571}{81} -
\frac{67\zeta_2}{36} + \frac{341\zeta_3}{18}
\nn
\\
&&
+ \frac{7\zeta_4}{8} + \ep\left(-\frac{7}{6} \zeta _3 \zeta _2-\frac{139 \zeta
   _2}{54}+\frac{2077 \zeta _3}{54}+\frac{2035
   \zeta _4}{48}-\frac{247 \zeta
   _5}{10}-\frac{3410}{243} \right) 
\nn
\\
&&
+ \ep^2 \left( -\frac{205 \zeta _3^2}{18}+\frac{341 \zeta _2
   \zeta _3}{18}+\frac{6388 \zeta
   _3}{81}-\frac{436 \zeta _2}{81}+\frac{12395
   \zeta _4}{144}+\frac{5621 \zeta
   _5}{30}
\right.
\nn
\\
&&
\left.
-\frac{3307 \zeta
   _6}{48}-\frac{20428}{729}
\right)\Bigg]
 + \mathcal{O}(\ep^3) \Bigg\}.
\label{eq:master}
\end{eqnarray}
Eq.~(\ref{eq:master}) is the main result of this paper. We remind the
reader that this result is for the unrenormalized amplitudes. To
obtain the
renormalized ones, one only needs to perform a renormalization on the
strong coupling $\als$. We have
checked Eq.~(\ref{eq:master}) against the two-loop splitting amplitudes in
the soft limit calculated in
Refs.~\cite{Bern:2004cz,Badger:2004uk}, and found full agreement
to order $\ep^0$. To the best of our knowledge, the order $\ep$ and
$\ep^2$ terms presented in this paper are new. 

\subsection{Single soft-gluon current in ${\cal N}=4$ Super-Yang-Mills theory}
\label{sec:4c}

Using the generic results presented above, it's straightforward to obtain the
single soft-gluon current in ${\cal N}=4$ Super-Yang-Mills theory, by setting $N_f=4C_A$, $N_s=6C_A$, and $D_s=4$~(corresponding
to FDH scheme~\cite{Bern:1991aq,Bern:2002zk}) in
Eq.~(\ref{eq:reduction}):
\begin{eqnarray}
  \label{eq:neq4}
   S^{(2)}_{12,{\cal N}=4}(q) = g_s^4 C^2_A \frac{p_1\cdot
      p_2}{(q\cdot p_1)(q\cdot p_2)}\left[ - \frac{1-4\ep}{3\ep}
   I_1 + \frac{1-2\ep}{\ep} I_2 + \frac{8}{3} I_3\right].
\end{eqnarray}
This result is remarkably simple. It becomes obvious that the result in
${\cal N}=4$ Super-Yang-Mills theory has uniform transcendentality, as
long as $I_3$ does. Substituting the explicit form of the master
integrals into Eq.~(\ref{eq:neq4}), we obtain
\begin{multline}
    S^{(2)}_{12,{\cal N}=4}(q) = S^{(0)}_{12}(q) \left(\frac{\als}{4\pi}
 \right)^2 S^2_\ep C^2_A
\\
\times\Bigg[ \frac{1}{2\ep^4} + \frac{\zeta_2}{\ep^2} -
\frac{11\zeta_3}{6\ep} + \frac{7\zeta_4}{8} + \ep \left( -
  \frac{7\zeta_2\zeta_3}{6} - \frac{247\zeta_5}{10} \right) 
+ \ep^2
\left( - \frac{205\zeta^2_3}{18} - \frac{3307\zeta_6}{48}\right)
+ \mathcal{O}(\ep^3) \Bigg].
 \label{eq:sym}
\end{multline}
We note that at leading transcendentality, the eikonal soft function in ${\cal N}=4$ Super-Yang-Mills theory
coincides with the one in QCD through $\ep^2$,
as also happens in some other context~\cite{Kotikov:2004er}.

It's also interesting to notice that Eq.~(\ref{eq:sym}) can be
reorganized as~\cite{Anastasiou:2003kj}~\footnote{We are grateful to
  Lance Dixon for pointing us to the discussion in the rest of this section.}
\begin{eqnarray}
      S^{(2)}_{12,{\cal N}=4}(q) &\equiv& 4S^{(0)}_{12}(q) \left(\frac{\als}{4\pi}
 \right)^2   S^2_\ep  C^2_A r_S^{(2)}(\ep)
\nn
\\
&=&
4S^{(0)}_{12}(q) \left(\frac{\als}{4\pi}
 \right)^2   S^2_\ep  C^2_A 
 \left( \frac{1}{2} \left(r_S^{(1)}(\ep)\right)^2 + f(\ep) r_S^{(1)}(2\ep) \right)
 + \mathcal{O}(\ep),
\label{eq:exp}
\end{eqnarray}
where $r_S^{(1)}(\ep) =-e^{\ep\gamma_E} \frac{
  \Gamma^3(1-\ep)\Gamma^2(1+\ep)}{2\ep^2\Gamma(1-2\ep)}$ is the soft
limit of the one-loop collinear splitting amplitudes in ${\cal N}=4$
Super-Yang-Mills theory~(up to an
overall $z$-dependent factor, same below), and $f(\ep) =
-\sum_{i=1}^\infty \zeta_{i+1} \ep^{i-1}$~\cite{Anastasiou:2003kj}. Eq.~(\ref{eq:exp}) makes explicit
the iterative structure of ${\cal N}=4$ splitting amplitudes and eikonal
function~\cite{Bern:2005iz}. Eq.~(\ref{eq:exp}) also determines the
soft limit of two-loop splitting amplitudes beyond order $\ep^0$,
\begin{eqnarray}
  r^{(2)}_S(\ep) &=& \frac{1}{8\ep^4} + \frac{\zeta_2}{4\ep^2} -
\frac{11\zeta_3}{24\ep} + \frac{7\zeta_4}{32} + \ep \left( -
  \frac{7\zeta_2\zeta_3}{24} - \frac{247\zeta_5}{40} \right) 
\nn
\\
&&+ \ep^2
\left( - \frac{205\zeta^2_3}{72} - \frac{3307\zeta_6}{192}\right) 
+\mathcal{O}(\ep^3).
\label{eq:r2}
\end{eqnarray}
At three loops, the soft limit of splitting amplitudes at leading color is predicted to be~\cite{Bern:2005iz}
\begin{eqnarray}
  r^{(3)}_S(\ep) = - \frac{1}{3} \left(r^{(1)}_S(\ep) \right)^3 +
  r^{(1)}_S(\ep) r^{(2)}_S(\ep) + f^{(3)}(\ep) r^{(1)}_S(3\ep) + \mathcal{O}(\ep),
\end{eqnarray}
where $f^{(3)}$ has been calculated through order
$\ep^2$~\cite{Spradlin:2008uu}, 
\begin{eqnarray}
  f^{(3)}(\ep) = \frac{11\zeta_4}{2} + (5\zeta_2\zeta_3 + 6\zeta_5)\ep
  + a\ep^2 + \mathcal{O}(\ep^3),
\end{eqnarray}
with $a=85.263\pm 0.004$. Using the above results, we obtain
\begin{eqnarray}
  r^{(3)}_S(\ep) = - \frac{1}{48\ep^6}  - \frac{3\zeta_2}{32\ep^4} +
  \frac{\zeta_3}{12\ep^3} - \frac{1487\zeta_4}{2304 \ep^2}  -
  \frac{13\zeta_2\zeta_3}{144\ep} + \frac{71\zeta_5}{30\ep} +
  \frac{11005\zeta_6}{2048} + \frac{167\zeta^2_3}{96} - \frac{a}{18}
  + \mathcal{O}(\ep).
\nn
\\
\label{eq:rr}
\end{eqnarray}
For completeness, the ${\cal N}=4$ eikonal function at leading color
is then given by
\begin{eqnarray}
      S^{(3)}_{12,{\cal N}=4}(q) &=& 8S^{(0)}_{12}(q) \left(\frac{\als }{4\pi}
 \right)^3  S^3_\ep  C^3_A r_S^{(3)}(\ep).
\end{eqnarray}
We note that Eq.~(\ref{eq:rr}) is actually exact through order $\ep^{-1}$ for finite $N_c$. There are
potential $\frac{1}{N_c}$ corrections, starting from order
$\ep^0$. Unlike the one-loop and two-loop cases, these corrections
would depend explicitly on the color representation of the hard
partons, through the product of fourth order invariant tensor,
$d^{ijkl}_Rd^{ijkl}_A$. An explicit calculation of these corrections
would be necessary in obtaining them.

Using the iterative predictions for the ${\cal N}=4$ splitting
amplitudes~\cite{Bern:2005iz} and the cusp anomalous dimension at leading
color~\cite{Bern:2006ew}, the results above can further determine the
leading-color ${\cal N}=4$ eikonal function at four loops through
order $\ep^{-2}$.

\section{C\lowercase{onclusion}}
\label{sec:5}

In this paper we have computed the single soft-gluon current to
two-loop order. we have compared our results with those in
Refs.~\cite{Bern:2004cz,Badger:2004uk}, and found full agreement to
order $\ep^0$. The order $\ep$ and order $\ep^2$ terms presented in
this paper are new. As a by-product, we have also given the soft-gluon
current in ${\cal N}=4$ Super-Yang-Mills theory to order $\ep^2$,
which in turn enables us to derive the splitting amplitudes in the soft limit,
or the single soft-gluon current, at three loops and large $N_c$, using the results of
Refs.~\cite{Bern:2005iz,Spradlin:2008uu}. We observe uniform
transcendentality for the single soft-gluon current in ${\cal N}=4$
Super-Yang-Mills theory, and confirm that the
leading transcendentality terms for the eikonal function are the same in QCD and ${\cal
  N}=4$ Super-Yang-Mills theory at two loops, up to order $\ep^2$. 

The main purpose of the computation done in this paper is to provide
the necessary ingredient for a calculation of Higgs production cross
section at hadron collider at NNNLO. A lot of progress have been made
recently in this
direction~\cite{Pak:2011hs,Anastasiou:2012kq,Hoschele:2012xc,Anastasiou:2013srw,Ball:2013bra,Buehler:2013fha}. A
useful step towards the full NNNLO QCD corrections is the soft-virtual
approximation at NNNLO. Using the results presented in this paper, the
cross section for Higgs $+$ one gluon emission can be computed by
trivially integrating over the soft-gluon phase space. The cross section
for Higgs $+$ 3 partons production in the soft limit has also been calculated recently in an
impressive paper~\cite{Anastasiou:2013srw}.  The only missing piece is
the cross section for Higgs $+$ 2 partons production at one loop in the soft
limit. It's reasonable to expect that the soft-virtual approximation
for Higgs production at hadron collider at NNNLO will be available in
the foreseeable future.

Besides Higgs physics, the soft gluon current at two loops is also
useful in soft-collinear-effective
theory~\cite{Bauer:2000ew,Bauer:2000yr,Bauer:2001yt,Beneke:2002ph}. For
example, the two-loop soft gluon current can be used to calculate soft
function at NNNLO. Finally, we have only computed this soft gluon
current with two energetic partons, or two Wilson lines. It's
certainly interesting to extend our results to processes with
an arbitrary number of Wilson lines. This will be relevant to jet physics
at NNNLO. It will also be useful in understanding the structure of
infrared divergences for multiple Wilson lines at three loops, see for example, Ref.~\cite{Gardi:2013ita}.

\begin{acknowledgements}
We are grateful to Lance Dixon for careful reading of the manuscript
and invaluable suggestions. 
We also thank Michael Peskin for useful discussion. This research is 
supported by the US Department of Energy under contract
DE–AC02–76SF00515. HXZ would like to thank Kavli Institute for
Theoretical Physics at UC Santa Barbara
and the Erwin Schr\"{o}dinger Institute at University of Vienna for
hospitality while part of this work was carried out.
While this paper was being completed, we learned of 
independent work~\cite{Duhr:2013msa} by Claude Duhr and Thomas Gehrmann
obtaining the soft-gluon current to all orders in $\epsilon$, which is
in full agreement with the Laurent expansions of the
soft-gluon current to order $\epsilon^2$ derived in this paper.

\end{acknowledgements}

\appendix*

\section{Evaluation of the master integral $I_3$ to order $\ep^2$}
\label{sec:6}

In this appendix, we briefly explain the evaluation of the most
difficult master integral, $I_3$. In fact, $I_1$ and $I_2$ can be obtained by
deleting two propagators from $I_3$. We proceed by first performing
the $k_2$ sub-loop integral by Feynman's trick,
\begin{eqnarray}
  I'_{3} &=& \int\! \frac{d^D k_2}{i\pi^{D/2}} 
\frac{1}{ [2(k_2+q)\cdot p_2]
  [k^2_2] [(k_1+k_2)^2] [(k_2+q)^2]}
\nn
\\
&=&
(-1)^{D/2} \Gamma\left(4 - \frac{D}{2}  \right)\int^{\infty}_0\! dy_2
\,\int^1_0 \! dx_1\, dx_2\, dx_3 \frac{ \delta(1-x_1-x_2-x_3)}{ \Delta^{4-D/2}},
\end{eqnarray}
where
\begin{eqnarray}
  \Delta = x_1 x_2 [k^2_1 ] +  x_1x_3 [(q+y_2 p_2)^2] + x_2 x_3 [(k_1 - q
  - y_2 p_2)^2].
\end{eqnarray}
The resulting Feynman parameter integral over $dx_i$ can be factorized
by introducing a two-fold Mellin-Barnes integral, 
\begin{eqnarray}
  \frac{1}{\Delta^{4-D/2}} &=& \int^{+i\infty}_{-i\infty} \!\frac{dz_1\,
    dz_2}{(2\pi i)^2} \Gamma(-z_1)\Gamma(-z_2)
\frac{  \Gamma\left(4-D/2 + z_1 + z_2 \right)}{\Gamma(4-D/2)} 
\nn
\\
&&
\times  \Big( x_1 x_2 [k^2_1 ]\Big)^{z_1} \Big(  x_1x_3 [(q+y_2
p_2)^2] \Big)^{z_2} \Big( x_2 x_3 [(k_1 - q
  - y_2 p_2)^2] \Big)^{-z_1 - z_2 - 4 + D/2},
\end{eqnarray}
where the contour for $z_i$ separates the poles of $\Gamma(\dots +
z_i)$ from those of $\Gamma(\dots - z_i)$. After this step, the
Feynman parameter integral over $dx_i$ can be done in closed form in
terms of $\Gamma$ functions. The remaining $k_1$ sub-loop integral has the
form
\begin{eqnarray}
  \int\! \frac{d^D k_1}{i\pi^{D/2}} \frac{1}{[k^2_1]^{1-z_1} [(k_1 - q
    - y_2 p_2)^2]^{4-D/2 + z_1 + z_2} [2 k_1 \cdot p_1] [ 2(q -
    k_1)\cdot p_2]},
\end{eqnarray}
which can be straightforwardly done. We then arrive at a two-fold
Mellins-Barnes integral representation for $I_3$,
\begin{eqnarray}
  I_3 &=& \frac{1}{8(16\pi^2)^2}S^2_\ep e^{2\ep\gamma_E}  \Gamma(5-D) 
\nn
\\
&&
\times
\int^{+i\infty}_{-i\infty} \!\frac{dz_1\,
    dz_2}{(2\pi i)^2}
\Gamma(-z_1) \Gamma(-z_2)
  \Gamma(z_2 + 1) \Gamma\left( \frac{D}{2} - z_1 - 2\right)  
\Gamma\left( \frac{D}{2} +  z_1 - 2\right) 
\nn
\\
&&
\times
\frac{\Gamma\left( \frac{D}{2} - z_2 - 3\right)
 \Gamma\left( -D  + z_2 + 6\right)  \Gamma\left(1 + z_1 + z_2 \right)
\Gamma( D - z_1 - z_2 - 5)
 }{ \Gamma\left( 1-z_1\right) \Gamma\left( 2 + z_2\right) 
\Gamma\left( \frac{3D}{2} - z_2 - 7\right) }.
\end{eqnarray}
We were not able to find an all order in $\ep$ solution of this
integral. Instead, we calculate the Laurent expansion of the
Mellin-Barnes integral to order $\ep^2$, which is relevant for NNNLO
phenomenology. To that end, we make use of the MATHEMATICA pachages
MB~\cite{Czakon:2005rk} and BARNESROUTINES of D.~Kosower to resolve the singularity,
to expand the integrand in $\ep$, and to apply the Barnes lemma in an
automatic way, which results in a series of one-fold and two-fold
Mellin-Barnes integrals. The one-fold integral can easily be done
numerically using MATHEMATICA's {\tt NIntegrate} routine, and the results
can be converted into Riemann zeta values using the PSLQ algorithm~\cite{pslq1,pslq2}. The
only remaining two-fold Mellin-Barnes integral is
\begin{eqnarray}
  &&\int^{+i\infty}_{-i\infty} \!\frac{dz_1\,
    dz_2}{(2\pi i)^2} \frac{\Gamma(-z_1)^2 \Gamma(z_1) \Gamma(-z_2)
    \Gamma(1+z_2) \Gamma(-1-z_1 -z_2) \Gamma(1+z_1 +
    z_2)}{\Gamma(1-z_1) }
\nn
\\
&&
\times
\left( \psi(-1-z_2) + \psi(2+z_2)\right) 
\left( 2\psi(-1-z_1-z_2) + \psi(-z_1) + \psi(z_1) \right),
\end{eqnarray}
where $\psi(x)$ is the logarithmic derivative of $\Gamma$ function, and
the integration contours are straight vertical lines defined by
\begin{eqnarray}
  \mathrm{Re}(z_1) = -\frac{1091}{1641},\qquad
\mathrm{Re}(z_2) = -\frac{554}{1671}.
\end{eqnarray}
The integral can be performed by closing the contour to the left or
right, and summing up the residues at the poles. The results are
double sums of the form
\begin{eqnarray}
  \sum^{\infty}_{m,n=1} \frac{S_{\vec{i}_1}(m)}{m^{j_1}}
  \frac{S_{\vec{i}_2}(n)}{n^{j_2}} \frac{S_{\vec{i}_3}(m+n)}{(m+n)^{j_3}},
\end{eqnarray}
where $S_{\vec{i}}(k)$ are nested harmonic sums defined in
Ref.~\cite{Vermaseren:1998uu}. The summation can be conveniently done
using XSummer~\cite{Moch:2005uc}. The final result for this master
integral is checked numerically using the package
FIESTA~\cite{Smirnov:2009pb} and the author's personal tool, based on the method of sector
decomposition~\cite{Binoth:2000ps}.

\bibliography{twoloopg}

\begin{thebibliography}{80}%
\makeatletter
\providecommand \@ifxundefined [1]{%
 \@ifx{#1\undefined}
}%
\providecommand \@ifnum [1]{%
 \ifnum #1\expandafter \@firstoftwo
 \else \expandafter \@secondoftwo
 \fi
}%
\providecommand \@ifx [1]{%
 \ifx #1\expandafter \@firstoftwo
 \else \expandafter \@secondoftwo
 \fi
}%
\providecommand \natexlab [1]{#1}%
\providecommand \enquote  [1]{``#1''}%
\providecommand \bibnamefont  [1]{#1}%
\providecommand \bibfnamefont [1]{#1}%
\providecommand \citenamefont [1]{#1}%
\providecommand \href@noop [0]{\@secondoftwo}%
\providecommand \href [0]{\begingroup \@sanitize@url \@href}%
\providecommand \@href[1]{\@@startlink{#1}\@@href}%
\providecommand \@@href[1]{\endgroup#1\@@endlink}%
\providecommand \@sanitize@url [0]{\catcode `\\12\catcode `\$12\catcode
  `\&12\catcode `\#12\catcode `\^12\catcode `\_12\catcode `\%12\relax}%
\providecommand \@@startlink[1]{}%
\providecommand \@@endlink[0]{}%
\providecommand \url  [0]{\begingroup\@sanitize@url \@url }%
\providecommand \@url [1]{\endgroup\@href {#1}{\urlprefix }}%
\providecommand \urlprefix  [0]{URL }%
\providecommand \Eprint [0]{\href }%
\providecommand \doibase [0]{http://dx.doi.org/}%
\providecommand \selectlanguage [0]{\@gobble}%
\providecommand \bibinfo  [0]{\@secondoftwo}%
\providecommand \bibfield  [0]{\@secondoftwo}%
\providecommand \translation [1]{[#1]}%
\providecommand \BibitemOpen [0]{}%
\providecommand \bibitemStop [0]{}%
\providecommand \bibitemNoStop [0]{.\EOS\space}%
\providecommand \EOS [0]{\spacefactor3000\relax}%
\providecommand \BibitemShut  [1]{\csname bibitem#1\endcsname}%
\let\auto@bib@innerbib\@empty
\bibitem [{\citenamefont {Altarelli}\ and\ \citenamefont
  {Parisi}(1977)}]{Altarelli:1977zs}%
  \BibitemOpen
  \bibfield  {author} {\bibinfo {author} {\bibfnamefont {G.}~\bibnamefont
  {Altarelli}}\ and\ \bibinfo {author} {\bibfnamefont {G.}~\bibnamefont
  {Parisi}},\ }\href {\doibase 10.1016/0550-3213(77)90384-4} {\bibfield
  {journal} {\bibinfo  {journal} {Nucl.Phys.}\ }\textbf {\bibinfo {volume}
  {B126}},\ \bibinfo {pages} {298} (\bibinfo {year} {1977})}\BibitemShut
  {NoStop}%
\bibitem [{\citenamefont {Bassetto}\ \emph {et~al.}(1983)\citenamefont
  {Bassetto}, \citenamefont {Ciafaloni},\ and\ \citenamefont
  {Marchesini}}]{Bassetto:1984ik}%
  \BibitemOpen
  \bibfield  {author} {\bibinfo {author} {\bibfnamefont {A.}~\bibnamefont
  {Bassetto}}, \bibinfo {author} {\bibfnamefont {M.}~\bibnamefont {Ciafaloni}},
  \ and\ \bibinfo {author} {\bibfnamefont {G.}~\bibnamefont {Marchesini}},\
  }\href {\doibase 10.1016/0370-1573(83)90083-2} {\bibfield  {journal}
  {\bibinfo  {journal} {Phys.Rept.}\ }\textbf {\bibinfo {volume} {100}},\
  \bibinfo {pages} {201} (\bibinfo {year} {1983})}\BibitemShut {NoStop}%
\bibitem [{\citenamefont {Berends}\ and\ \citenamefont
  {Giele}(1989)}]{Berends:1988zn}%
  \BibitemOpen
  \bibfield  {author} {\bibinfo {author} {\bibfnamefont {F.~A.}\ \bibnamefont
  {Berends}}\ and\ \bibinfo {author} {\bibfnamefont {W.}~\bibnamefont
  {Giele}},\ }\href {\doibase 10.1016/0550-3213(89)90398-2} {\bibfield
  {journal} {\bibinfo  {journal} {Nucl.Phys.}\ }\textbf {\bibinfo {volume}
  {B313}},\ \bibinfo {pages} {595} (\bibinfo {year} {1989})}\BibitemShut
  {NoStop}%
\bibitem [{\citenamefont {Dokshitzer}\ \emph {et~al.}(1991)\citenamefont
  {Dokshitzer}, \citenamefont {Khoze}, \citenamefont {Mueller},\ and\
  \citenamefont {Troian}}]{Dokshitzer:1991wu}%
  \BibitemOpen
  \bibfield  {author} {\bibinfo {author} {\bibfnamefont {Y.~L.}\ \bibnamefont
  {Dokshitzer}}, \bibinfo {author} {\bibfnamefont {V.~A.}\ \bibnamefont
  {Khoze}}, \bibinfo {author} {\bibfnamefont {A.~H.}\ \bibnamefont {Mueller}},
  \ and\ \bibinfo {author} {\bibfnamefont {S.}~\bibnamefont {Troian}},\
  }\href@noop {} {\bibfield  {journal} {\bibinfo  {journal} {\emph{Basics of
  perturbative QCD}, Editions Fronti\`{e}res, Gif-sur-Yvette}\ } (\bibinfo
  {year} {1991})}\BibitemShut {NoStop}%
\bibitem [{\citenamefont {Ellis}\ \emph {et~al.}(1996)\citenamefont {Ellis},
  \citenamefont {Stirling},\ and\ \citenamefont {Webber}}]{Ellis:1991qj}%
  \BibitemOpen
  \bibfield  {author} {\bibinfo {author} {\bibfnamefont {R.~K.}\ \bibnamefont
  {Ellis}}, \bibinfo {author} {\bibfnamefont {W.~J.}\ \bibnamefont {Stirling}},
  \ and\ \bibinfo {author} {\bibfnamefont {B.}~\bibnamefont {Webber}},\
  }\href@noop {} {\bibfield  {journal} {\bibinfo  {journal} {\emph{QCD and
  Collider Physics}, Cambridge University Press, Cambridge}\ } (\bibinfo {year}
  {1996})}\BibitemShut {NoStop}%
\bibitem [{\citenamefont {Campbell}\ and\ \citenamefont
  {Glover}(1998)}]{Campbell:1997hg}%
  \BibitemOpen
  \bibfield  {author} {\bibinfo {author} {\bibfnamefont {J.~M.}\ \bibnamefont
  {Campbell}}\ and\ \bibinfo {author} {\bibfnamefont {E.~W.~N.}\ \bibnamefont
  {Glover}},\ }\href {\doibase 10.1016/S0550-3213(98)00295-8} {\bibfield
  {journal} {\bibinfo  {journal} {Nucl.Phys.}\ }\textbf {\bibinfo {volume}
  {B527}},\ \bibinfo {pages} {264} (\bibinfo {year} {1998})},\ \Eprint
  {http://arxiv.org/abs/hep-ph/9710255} {arXiv:hep-ph/9710255 [hep-ph]}
  \BibitemShut {NoStop}%
\bibitem [{\citenamefont {Catani}\ and\ \citenamefont
  {Grazzini}(2000{\natexlab{a}})}]{Catani:1999ss}%
  \BibitemOpen
  \bibfield  {author} {\bibinfo {author} {\bibfnamefont {S.}~\bibnamefont
  {Catani}}\ and\ \bibinfo {author} {\bibfnamefont {M.}~\bibnamefont
  {Grazzini}},\ }\href {\doibase 10.1016/S0550-3213(99)00778-6} {\bibfield
  {journal} {\bibinfo  {journal} {Nucl.Phys.}\ }\textbf {\bibinfo {volume}
  {B570}},\ \bibinfo {pages} {287} (\bibinfo {year} {2000}{\natexlab{a}})},\
  \Eprint {http://arxiv.org/abs/hep-ph/9908523} {arXiv:hep-ph/9908523 [hep-ph]}
  \BibitemShut {NoStop}%
\bibitem [{\citenamefont {Bern}\ and\ \citenamefont
  {Chalmers}(1995)}]{Bern:1995ix}%
  \BibitemOpen
  \bibfield  {author} {\bibinfo {author} {\bibfnamefont {Z.}~\bibnamefont
  {Bern}}\ and\ \bibinfo {author} {\bibfnamefont {G.}~\bibnamefont
  {Chalmers}},\ }\href {\doibase 10.1016/0550-3213(95)00226-I} {\bibfield
  {journal} {\bibinfo  {journal} {Nucl.Phys.}\ }\textbf {\bibinfo {volume}
  {B447}},\ \bibinfo {pages} {465} (\bibinfo {year} {1995})},\ \Eprint
  {http://arxiv.org/abs/hep-ph/9503236} {arXiv:hep-ph/9503236 [hep-ph]}
  \BibitemShut {NoStop}%
\bibitem [{\citenamefont {Kosower}(1999)}]{Kosower:1999xi}%
  \BibitemOpen
  \bibfield  {author} {\bibinfo {author} {\bibfnamefont {D.~A.}\ \bibnamefont
  {Kosower}},\ }\href {\doibase 10.1016/S0550-3213(99)00251-5} {\bibfield
  {journal} {\bibinfo  {journal} {Nucl.Phys.}\ }\textbf {\bibinfo {volume}
  {B552}},\ \bibinfo {pages} {319} (\bibinfo {year} {1999})},\ \Eprint
  {http://arxiv.org/abs/hep-ph/9901201} {arXiv:hep-ph/9901201 [hep-ph]}
  \BibitemShut {NoStop}%
\bibitem [{\citenamefont {Catani}\ and\ \citenamefont
  {Grazzini}(1999)}]{Catani:1998nv}%
  \BibitemOpen
  \bibfield  {author} {\bibinfo {author} {\bibfnamefont {S.}~\bibnamefont
  {Catani}}\ and\ \bibinfo {author} {\bibfnamefont {M.}~\bibnamefont
  {Grazzini}},\ }\href {\doibase 10.1016/S0370-2693(98)01513-5} {\bibfield
  {journal} {\bibinfo  {journal} {Phys.Lett.}\ }\textbf {\bibinfo {volume}
  {B446}},\ \bibinfo {pages} {143} (\bibinfo {year} {1999})},\ \Eprint
  {http://arxiv.org/abs/hep-ph/9810389} {arXiv:hep-ph/9810389 [hep-ph]}
  \BibitemShut {NoStop}%
\bibitem [{\citenamefont {Del~Duca}\ \emph {et~al.}(2000)\citenamefont
  {Del~Duca}, \citenamefont {Frizzo},\ and\ \citenamefont
  {Maltoni}}]{DelDuca:1999ha}%
  \BibitemOpen
  \bibfield  {author} {\bibinfo {author} {\bibfnamefont {V.}~\bibnamefont
  {Del~Duca}}, \bibinfo {author} {\bibfnamefont {A.}~\bibnamefont {Frizzo}}, \
  and\ \bibinfo {author} {\bibfnamefont {F.}~\bibnamefont {Maltoni}},\ }\href
  {\doibase 10.1016/S0550-3213(99)00657-4} {\bibfield  {journal} {\bibinfo
  {journal} {Nucl.Phys.}\ }\textbf {\bibinfo {volume} {B568}},\ \bibinfo
  {pages} {211} (\bibinfo {year} {2000})},\ \Eprint
  {http://arxiv.org/abs/hep-ph/9909464} {arXiv:hep-ph/9909464 [hep-ph]}
  \BibitemShut {NoStop}%
\bibitem [{\citenamefont {Bern}\ \emph {et~al.}(1998)\citenamefont {Bern},
  \citenamefont {Del~Duca},\ and\ \citenamefont {Schmidt}}]{Bern:1998sc}%
  \BibitemOpen
  \bibfield  {author} {\bibinfo {author} {\bibfnamefont {Z.}~\bibnamefont
  {Bern}}, \bibinfo {author} {\bibfnamefont {V.}~\bibnamefont {Del~Duca}}, \
  and\ \bibinfo {author} {\bibfnamefont {C.~R.}\ \bibnamefont {Schmidt}},\
  }\href {\doibase 10.1016/S0370-2693(98)01495-6} {\bibfield  {journal}
  {\bibinfo  {journal} {Phys.Lett.}\ }\textbf {\bibinfo {volume} {B445}},\
  \bibinfo {pages} {168} (\bibinfo {year} {1998})},\ \Eprint
  {http://arxiv.org/abs/hep-ph/9810409} {arXiv:hep-ph/9810409 [hep-ph]}
  \BibitemShut {NoStop}%
\bibitem [{\citenamefont {Catani}(1998)}]{Catani:1998bh}%
  \BibitemOpen
  \bibfield  {author} {\bibinfo {author} {\bibfnamefont {S.}~\bibnamefont
  {Catani}},\ }\href {\doibase 10.1016/S0370-2693(98)00332-3} {\bibfield
  {journal} {\bibinfo  {journal} {Phys.Lett.}\ }\textbf {\bibinfo {volume}
  {B427}},\ \bibinfo {pages} {161} (\bibinfo {year} {1998})},\ \Eprint
  {http://arxiv.org/abs/hep-ph/9802439} {arXiv:hep-ph/9802439 [hep-ph]}
  \BibitemShut {NoStop}%
\bibitem [{\citenamefont {Bern}\ \emph {et~al.}(1999)\citenamefont {Bern},
  \citenamefont {Del~Duca}, \citenamefont {Kilgore},\ and\ \citenamefont
  {Schmidt}}]{Bern:1999ry}%
  \BibitemOpen
  \bibfield  {author} {\bibinfo {author} {\bibfnamefont {Z.}~\bibnamefont
  {Bern}}, \bibinfo {author} {\bibfnamefont {V.}~\bibnamefont {Del~Duca}},
  \bibinfo {author} {\bibfnamefont {W.~B.}\ \bibnamefont {Kilgore}}, \ and\
  \bibinfo {author} {\bibfnamefont {C.~R.}\ \bibnamefont {Schmidt}},\ }\href
  {\doibase 10.1103/PhysRevD.60.116001} {\bibfield  {journal} {\bibinfo
  {journal} {Phys.Rev.}\ }\textbf {\bibinfo {volume} {D60}},\ \bibinfo {pages}
  {116001} (\bibinfo {year} {1999})},\ \Eprint
  {http://arxiv.org/abs/hep-ph/9903516} {arXiv:hep-ph/9903516 [hep-ph]}
  \BibitemShut {NoStop}%
\bibitem [{\citenamefont {Catani}\ and\ \citenamefont
  {Grazzini}(2000{\natexlab{b}})}]{Catani:2000pi}%
  \BibitemOpen
  \bibfield  {author} {\bibinfo {author} {\bibfnamefont {S.}~\bibnamefont
  {Catani}}\ and\ \bibinfo {author} {\bibfnamefont {M.}~\bibnamefont
  {Grazzini}},\ }\href {\doibase 10.1016/S0550-3213(00)00572-1} {\bibfield
  {journal} {\bibinfo  {journal} {Nucl.Phys.}\ }\textbf {\bibinfo {volume}
  {B591}},\ \bibinfo {pages} {435} (\bibinfo {year} {2000}{\natexlab{b}})},\
  \Eprint {http://arxiv.org/abs/hep-ph/0007142} {arXiv:hep-ph/0007142 [hep-ph]}
  \BibitemShut {NoStop}%
\bibitem [{\citenamefont {Kosower}(2003)}]{Kosower:2003cz}%
  \BibitemOpen
  \bibfield  {author} {\bibinfo {author} {\bibfnamefont {D.~A.}\ \bibnamefont
  {Kosower}},\ }\href {\doibase 10.1103/PhysRevLett.91.061602} {\bibfield
  {journal} {\bibinfo  {journal} {Phys.Rev.Lett.}\ }\textbf {\bibinfo {volume}
  {91}},\ \bibinfo {pages} {061602} (\bibinfo {year} {2003})},\ \Eprint
  {http://arxiv.org/abs/hep-ph/0301069} {arXiv:hep-ph/0301069 [hep-ph]}
  \BibitemShut {NoStop}%
\bibitem [{\citenamefont {Catani}\ \emph {et~al.}(2004)\citenamefont {Catani},
  \citenamefont {de~Florian},\ and\ \citenamefont {Rodrigo}}]{Catani:2003vu}%
  \BibitemOpen
  \bibfield  {author} {\bibinfo {author} {\bibfnamefont {S.}~\bibnamefont
  {Catani}}, \bibinfo {author} {\bibfnamefont {D.}~\bibnamefont {de~Florian}},
  \ and\ \bibinfo {author} {\bibfnamefont {G.}~\bibnamefont {Rodrigo}},\ }\href
  {\doibase 10.1016/j.physletb.2004.02.039} {\bibfield  {journal} {\bibinfo
  {journal} {Phys.Lett.}\ }\textbf {\bibinfo {volume} {B586}},\ \bibinfo
  {pages} {323} (\bibinfo {year} {2004})},\ \Eprint
  {http://arxiv.org/abs/hep-ph/0312067} {arXiv:hep-ph/0312067 [hep-ph]}
  \BibitemShut {NoStop}%
\bibitem [{\citenamefont {Bern}\ \emph {et~al.}(2004)\citenamefont {Bern},
  \citenamefont {Dixon},\ and\ \citenamefont {Kosower}}]{Bern:2004cz}%
  \BibitemOpen
  \bibfield  {author} {\bibinfo {author} {\bibfnamefont {Z.}~\bibnamefont
  {Bern}}, \bibinfo {author} {\bibfnamefont {L.~J.}\ \bibnamefont {Dixon}}, \
  and\ \bibinfo {author} {\bibfnamefont {D.~A.}\ \bibnamefont {Kosower}},\
  }\href {\doibase 10.1088/1126-6708/2004/08/012} {\bibfield  {journal}
  {\bibinfo  {journal} {JHEP}\ }\textbf {\bibinfo {volume} {0408}},\ \bibinfo
  {pages} {012} (\bibinfo {year} {2004})},\ \Eprint
  {http://arxiv.org/abs/hep-ph/0404293} {arXiv:hep-ph/0404293 [hep-ph]}
  \BibitemShut {NoStop}%
\bibitem [{\citenamefont {Badger}\ and\ \citenamefont
  {Glover}(2004)}]{Badger:2004uk}%
  \BibitemOpen
  \bibfield  {author} {\bibinfo {author} {\bibfnamefont {S.}~\bibnamefont
  {Badger}}\ and\ \bibinfo {author} {\bibfnamefont {E.~W.~N.}\ \bibnamefont
  {Glover}},\ }\href {\doibase 10.1088/1126-6708/2004/07/040} {\bibfield
  {journal} {\bibinfo  {journal} {JHEP}\ }\textbf {\bibinfo {volume} {0407}},\
  \bibinfo {pages} {040} (\bibinfo {year} {2004})},\ \Eprint
  {http://arxiv.org/abs/hep-ph/0405236} {arXiv:hep-ph/0405236 [hep-ph]}
  \BibitemShut {NoStop}%
\bibitem [{\citenamefont {Gehrmann-De~Ridder}\ \emph
  {et~al.}(2007)\citenamefont {Gehrmann-De~Ridder}, \citenamefont {Gehrmann},
  \citenamefont {Glover},\ and\ \citenamefont
  {Heinrich}}]{GehrmannDeRidder:2007jk}%
  \BibitemOpen
  \bibfield  {author} {\bibinfo {author} {\bibfnamefont {A.}~\bibnamefont
  {Gehrmann-De~Ridder}}, \bibinfo {author} {\bibfnamefont {T.}~\bibnamefont
  {Gehrmann}}, \bibinfo {author} {\bibfnamefont {E.~W.~N.}\ \bibnamefont
  {Glover}}, \ and\ \bibinfo {author} {\bibfnamefont {G.}~\bibnamefont
  {Heinrich}},\ }\href {\doibase 10.1088/1126-6708/2007/11/058} {\bibfield
  {journal} {\bibinfo  {journal} {JHEP}\ }\textbf {\bibinfo {volume} {0711}},\
  \bibinfo {pages} {058} (\bibinfo {year} {2007})},\ \Eprint
  {http://arxiv.org/abs/0710.0346} {arXiv:0710.0346 [hep-ph]} \BibitemShut
  {NoStop}%
\bibitem [{\citenamefont {Bierenbaum}\ \emph {et~al.}(2012)\citenamefont
  {Bierenbaum}, \citenamefont {Czakon},\ and\ \citenamefont
  {Mitov}}]{Bierenbaum:2011gg}%
  \BibitemOpen
  \bibfield  {author} {\bibinfo {author} {\bibfnamefont {I.}~\bibnamefont
  {Bierenbaum}}, \bibinfo {author} {\bibfnamefont {M.}~\bibnamefont {Czakon}},
  \ and\ \bibinfo {author} {\bibfnamefont {A.}~\bibnamefont {Mitov}},\ }\href
  {\doibase 10.1016/j.nuclphysb.2011.11.002} {\bibfield  {journal} {\bibinfo
  {journal} {Nucl.Phys.}\ }\textbf {\bibinfo {volume} {B856}},\ \bibinfo
  {pages} {228} (\bibinfo {year} {2012})},\ \Eprint
  {http://arxiv.org/abs/1107.4384} {arXiv:1107.4384 [hep-ph]} \BibitemShut
  {NoStop}%
\bibitem [{\citenamefont {Catani}\ \emph {et~al.}(2012)\citenamefont {Catani},
  \citenamefont {de~Florian},\ and\ \citenamefont {Rodrigo}}]{Catani:2011st}%
  \BibitemOpen
  \bibfield  {author} {\bibinfo {author} {\bibfnamefont {S.}~\bibnamefont
  {Catani}}, \bibinfo {author} {\bibfnamefont {D.}~\bibnamefont {de~Florian}},
  \ and\ \bibinfo {author} {\bibfnamefont {G.}~\bibnamefont {Rodrigo}},\ }\href
  {\doibase 10.1007/JHEP07(2012)026} {\bibfield  {journal} {\bibinfo  {journal}
  {JHEP}\ }\textbf {\bibinfo {volume} {1207}},\ \bibinfo {pages} {026}
  (\bibinfo {year} {2012})},\ \Eprint {http://arxiv.org/abs/1112.4405}
  {arXiv:1112.4405 [hep-ph]} \BibitemShut {NoStop}%
\bibitem [{\citenamefont {Currie}\ \emph {et~al.}(2013)\citenamefont {Currie},
  \citenamefont {Glover},\ and\ \citenamefont {Wells}}]{Currie:2013vh}%
  \BibitemOpen
  \bibfield  {author} {\bibinfo {author} {\bibfnamefont {J.}~\bibnamefont
  {Currie}}, \bibinfo {author} {\bibfnamefont {E.~W.~N.}\ \bibnamefont
  {Glover}}, \ and\ \bibinfo {author} {\bibfnamefont {S.}~\bibnamefont
  {Wells}},\ }\href {\doibase 10.1007/JHEP04(2013)066} {\bibfield  {journal}
  {\bibinfo  {journal} {JHEP}\ }\textbf {\bibinfo {volume} {1304}},\ \bibinfo
  {pages} {066} (\bibinfo {year} {2013})},\ \Eprint
  {http://arxiv.org/abs/1301.4693} {arXiv:1301.4693 [hep-ph]} \BibitemShut
  {NoStop}%
\bibitem [{\citenamefont {Feige}\ and\ \citenamefont
  {Schwartz}(2013)}]{Feige:2013zla}%
  \BibitemOpen
  \bibfield  {author} {\bibinfo {author} {\bibfnamefont {I.}~\bibnamefont
  {Feige}}\ and\ \bibinfo {author} {\bibfnamefont {M.~D.}\ \bibnamefont
  {Schwartz}},\ }\href@noop {} {\  (\bibinfo {year} {2013})},\ \Eprint
  {http://arxiv.org/abs/1306.6341} {arXiv:1306.6341 [hep-th]} \BibitemShut
  {NoStop}%
\bibitem [{\citenamefont {Gehrmann-De~Ridder}\ \emph
  {et~al.}(2005)\citenamefont {Gehrmann-De~Ridder}, \citenamefont {Gehrmann},\
  and\ \citenamefont {Glover}}]{GehrmannDeRidder:2005cm}%
  \BibitemOpen
  \bibfield  {author} {\bibinfo {author} {\bibfnamefont {A.}~\bibnamefont
  {Gehrmann-De~Ridder}}, \bibinfo {author} {\bibfnamefont {T.}~\bibnamefont
  {Gehrmann}}, \ and\ \bibinfo {author} {\bibfnamefont {E.~W.~N.}\ \bibnamefont
  {Glover}},\ }\href {\doibase 10.1088/1126-6708/2005/09/056} {\bibfield
  {journal} {\bibinfo  {journal} {JHEP}\ }\textbf {\bibinfo {volume} {0509}},\
  \bibinfo {pages} {056} (\bibinfo {year} {2005})},\ \Eprint
  {http://arxiv.org/abs/hep-ph/0505111} {arXiv:hep-ph/0505111 [hep-ph]}
  \BibitemShut {NoStop}%
\bibitem [{\citenamefont {Somogyi}\ \emph {et~al.}(2005)\citenamefont
  {Somogyi}, \citenamefont {Trocsanyi},\ and\ \citenamefont
  {Del~Duca}}]{Somogyi:2005xz}%
  \BibitemOpen
  \bibfield  {author} {\bibinfo {author} {\bibfnamefont {G.}~\bibnamefont
  {Somogyi}}, \bibinfo {author} {\bibfnamefont {Z.}~\bibnamefont {Trocsanyi}},
  \ and\ \bibinfo {author} {\bibfnamefont {V.}~\bibnamefont {Del~Duca}},\
  }\href {\doibase 10.1088/1126-6708/2005/06/024} {\bibfield  {journal}
  {\bibinfo  {journal} {JHEP}\ }\textbf {\bibinfo {volume} {0506}},\ \bibinfo
  {pages} {024} (\bibinfo {year} {2005})},\ \Eprint
  {http://arxiv.org/abs/hep-ph/0502226} {arXiv:hep-ph/0502226 [hep-ph]}
  \BibitemShut {NoStop}%
\bibitem [{\citenamefont {Czakon}(2010)}]{Czakon:2010td}%
  \BibitemOpen
  \bibfield  {author} {\bibinfo {author} {\bibfnamefont {M.}~\bibnamefont
  {Czakon}},\ }\href {\doibase 10.1016/j.physletb.2010.08.036} {\bibfield
  {journal} {\bibinfo  {journal} {Phys.Lett.}\ }\textbf {\bibinfo {volume}
  {B693}},\ \bibinfo {pages} {259} (\bibinfo {year} {2010})},\ \Eprint
  {http://arxiv.org/abs/1005.0274} {arXiv:1005.0274 [hep-ph]} \BibitemShut
  {NoStop}%
\bibitem [{\citenamefont {Baernreuther}\ \emph {et~al.}(2012)\citenamefont
  {Baernreuther}, \citenamefont {Czakon},\ and\ \citenamefont
  {Mitov}}]{Baernreuther:2012ws}%
  \BibitemOpen
  \bibfield  {author} {\bibinfo {author} {\bibfnamefont {P.}~\bibnamefont
  {Baernreuther}}, \bibinfo {author} {\bibfnamefont {M.}~\bibnamefont
  {Czakon}}, \ and\ \bibinfo {author} {\bibfnamefont {A.}~\bibnamefont
  {Mitov}},\ }\href {\doibase 10.1103/PhysRevLett.109.132001} {\bibfield
  {journal} {\bibinfo  {journal} {Phys.Rev.Lett.}\ }\textbf {\bibinfo {volume}
  {109}},\ \bibinfo {pages} {132001} (\bibinfo {year} {2012})},\ \Eprint
  {http://arxiv.org/abs/1204.5201} {arXiv:1204.5201 [hep-ph]} \BibitemShut
  {NoStop}%
\bibitem [{\citenamefont {Czakon}\ \emph {et~al.}(2013)\citenamefont {Czakon},
  \citenamefont {Fiedler},\ and\ \citenamefont {Mitov}}]{Czakon:2013goa}%
  \BibitemOpen
  \bibfield  {author} {\bibinfo {author} {\bibfnamefont {M.}~\bibnamefont
  {Czakon}}, \bibinfo {author} {\bibfnamefont {P.}~\bibnamefont {Fiedler}}, \
  and\ \bibinfo {author} {\bibfnamefont {A.}~\bibnamefont {Mitov}},\ }\href
  {\doibase 10.1103/PhysRevLett.110.252004} {\bibfield  {journal} {\bibinfo
  {journal} {Phys.Rev.Lett.}\ }\textbf {\bibinfo {volume} {110}},\ \bibinfo
  {pages} {252004} (\bibinfo {year} {2013})},\ \Eprint
  {http://arxiv.org/abs/1303.6254} {arXiv:1303.6254 [hep-ph]} \BibitemShut
  {NoStop}%
\bibitem [{\citenamefont {Ridder}\ \emph {et~al.}(2013)\citenamefont {Ridder},
  \citenamefont {Gehrmann}, \citenamefont {Glover},\ and\ \citenamefont
  {Pires}}]{Ridder:2013mf}%
  \BibitemOpen
  \bibfield  {author} {\bibinfo {author} {\bibfnamefont {A.~G.-D.}\
  \bibnamefont {Ridder}}, \bibinfo {author} {\bibfnamefont {T.}~\bibnamefont
  {Gehrmann}}, \bibinfo {author} {\bibfnamefont {E.~W.~N.}\ \bibnamefont
  {Glover}}, \ and\ \bibinfo {author} {\bibfnamefont {J.}~\bibnamefont
  {Pires}},\ }\href {\doibase 10.1103/PhysRevLett.110.162003} {\bibfield
  {journal} {\bibinfo  {journal} {Phys.Rev.Lett.}\ }\textbf {\bibinfo {volume}
  {110}},\ \bibinfo {pages} {162003} (\bibinfo {year} {2013})},\ \Eprint
  {http://arxiv.org/abs/1301.7310} {arXiv:1301.7310 [hep-ph]} \BibitemShut
  {NoStop}%
\bibitem [{\citenamefont {Boughezal}\ \emph {et~al.}(2013)\citenamefont
  {Boughezal}, \citenamefont {Caola}, \citenamefont {Melnikov}, \citenamefont
  {Petriello},\ and\ \citenamefont {Schulze}}]{Boughezal:2013uia}%
  \BibitemOpen
  \bibfield  {author} {\bibinfo {author} {\bibfnamefont {R.}~\bibnamefont
  {Boughezal}}, \bibinfo {author} {\bibfnamefont {F.}~\bibnamefont {Caola}},
  \bibinfo {author} {\bibfnamefont {K.}~\bibnamefont {Melnikov}}, \bibinfo
  {author} {\bibfnamefont {F.}~\bibnamefont {Petriello}}, \ and\ \bibinfo
  {author} {\bibfnamefont {M.}~\bibnamefont {Schulze}},\ }\href {\doibase
  10.1007/JHEP06(2013)072} {\bibfield  {journal} {\bibinfo  {journal} {JHEP}\
  }\textbf {\bibinfo {volume} {1306}},\ \bibinfo {pages} {072} (\bibinfo {year}
  {2013})},\ \Eprint {http://arxiv.org/abs/1302.6216} {arXiv:1302.6216
  [hep-ph]} \BibitemShut {NoStop}%
\bibitem [{\citenamefont {Aad}\ \emph {et~al.}(2012)\citenamefont {Aad} \emph
  {et~al.}}]{Aad:2012tfa}%
  \BibitemOpen
  \bibfield  {author} {\bibinfo {author} {\bibfnamefont {G.}~\bibnamefont
  {Aad}} \emph {et~al.} (\bibinfo {collaboration} {ATLAS Collaboration}),\
  }\href {\doibase 10.1016/j.physletb.2012.08.020} {\bibfield  {journal}
  {\bibinfo  {journal} {Phys.Lett.}\ }\textbf {\bibinfo {volume} {B716}},\
  \bibinfo {pages} {1} (\bibinfo {year} {2012})},\ \Eprint
  {http://arxiv.org/abs/1207.7214} {arXiv:1207.7214 [hep-ex]} \BibitemShut
  {NoStop}%
\bibitem [{\citenamefont {Chatrchyan}\ \emph {et~al.}(2012)\citenamefont
  {Chatrchyan} \emph {et~al.}}]{Chatrchyan:2012ufa}%
  \BibitemOpen
  \bibfield  {author} {\bibinfo {author} {\bibfnamefont {S.}~\bibnamefont
  {Chatrchyan}} \emph {et~al.} (\bibinfo {collaboration} {CMS Collaboration}),\
  }\href {\doibase 10.1016/j.physletb.2012.08.021} {\bibfield  {journal}
  {\bibinfo  {journal} {Phys.Lett.}\ }\textbf {\bibinfo {volume} {B716}},\
  \bibinfo {pages} {30} (\bibinfo {year} {2012})},\ \Eprint
  {http://arxiv.org/abs/1207.7235} {arXiv:1207.7235 [hep-ex]} \BibitemShut
  {NoStop}%
\bibitem [{\citenamefont {Harlander}\ and\ \citenamefont
  {Kilgore}(2002)}]{Harlander:2002wh}%
  \BibitemOpen
  \bibfield  {author} {\bibinfo {author} {\bibfnamefont {R.~V.}\ \bibnamefont
  {Harlander}}\ and\ \bibinfo {author} {\bibfnamefont {W.~B.}\ \bibnamefont
  {Kilgore}},\ }\href {\doibase 10.1103/PhysRevLett.88.201801} {\bibfield
  {journal} {\bibinfo  {journal} {Phys.Rev.Lett.}\ }\textbf {\bibinfo {volume}
  {88}},\ \bibinfo {pages} {201801} (\bibinfo {year} {2002})},\ \Eprint
  {http://arxiv.org/abs/hep-ph/0201206} {arXiv:hep-ph/0201206 [hep-ph]}
  \BibitemShut {NoStop}%
\bibitem [{\citenamefont {Anastasiou}\ and\ \citenamefont
  {Melnikov}(2002)}]{Anastasiou:2002yz}%
  \BibitemOpen
  \bibfield  {author} {\bibinfo {author} {\bibfnamefont {C.}~\bibnamefont
  {Anastasiou}}\ and\ \bibinfo {author} {\bibfnamefont {K.}~\bibnamefont
  {Melnikov}},\ }\href {\doibase 10.1016/S0550-3213(02)00837-4} {\bibfield
  {journal} {\bibinfo  {journal} {Nucl.Phys.}\ }\textbf {\bibinfo {volume}
  {B646}},\ \bibinfo {pages} {220} (\bibinfo {year} {2002})},\ \Eprint
  {http://arxiv.org/abs/hep-ph/0207004} {arXiv:hep-ph/0207004 [hep-ph]}
  \BibitemShut {NoStop}%
\bibitem [{\citenamefont {Ravindran}\ \emph {et~al.}(2003)\citenamefont
  {Ravindran}, \citenamefont {Smith},\ and\ \citenamefont {van
  Neerven}}]{Ravindran:2003um}%
  \BibitemOpen
  \bibfield  {author} {\bibinfo {author} {\bibfnamefont {V.}~\bibnamefont
  {Ravindran}}, \bibinfo {author} {\bibfnamefont {J.}~\bibnamefont {Smith}}, \
  and\ \bibinfo {author} {\bibfnamefont {W.~L.}\ \bibnamefont {van Neerven}},\
  }\href {\doibase 10.1016/S0550-3213(03)00457-7} {\bibfield  {journal}
  {\bibinfo  {journal} {Nucl.Phys.}\ }\textbf {\bibinfo {volume} {B665}},\
  \bibinfo {pages} {325} (\bibinfo {year} {2003})},\ \Eprint
  {http://arxiv.org/abs/hep-ph/0302135} {arXiv:hep-ph/0302135 [hep-ph]}
  \BibitemShut {NoStop}%
\bibitem [{\citenamefont {Catani}\ and\ \citenamefont
  {Grazzini}(2007)}]{Catani:2007vq}%
  \BibitemOpen
  \bibfield  {author} {\bibinfo {author} {\bibfnamefont {S.}~\bibnamefont
  {Catani}}\ and\ \bibinfo {author} {\bibfnamefont {M.}~\bibnamefont
  {Grazzini}},\ }\href {\doibase 10.1103/PhysRevLett.98.222002} {\bibfield
  {journal} {\bibinfo  {journal} {Phys.Rev.Lett.}\ }\textbf {\bibinfo {volume}
  {98}},\ \bibinfo {pages} {222002} (\bibinfo {year} {2007})},\ \Eprint
  {http://arxiv.org/abs/hep-ph/0703012} {arXiv:hep-ph/0703012 [hep-ph]}
  \BibitemShut {NoStop}%
\bibitem [{\citenamefont {Dittmaier}\ \emph {et~al.}(2011)\citenamefont
  {Dittmaier} \emph {et~al.}}]{Dittmaier:2011ti}%
  \BibitemOpen
  \bibfield  {author} {\bibinfo {author} {\bibfnamefont {S.}~\bibnamefont
  {Dittmaier}} \emph {et~al.} (\bibinfo {collaboration} {LHC Higgs Cross
  Section Working Group}),\ }\href@noop {} {\  (\bibinfo {year} {2011})},\
  \Eprint {http://arxiv.org/abs/1101.0593} {arXiv:1101.0593 [hep-ph]}
  \BibitemShut {NoStop}%
\bibitem [{\citenamefont {Boughezal}\ \emph {et~al.}(2012)\citenamefont
  {Boughezal}, \citenamefont {Melnikov},\ and\ \citenamefont
  {Petriello}}]{Boughezal:2011jf}%
  \BibitemOpen
  \bibfield  {author} {\bibinfo {author} {\bibfnamefont {R.}~\bibnamefont
  {Boughezal}}, \bibinfo {author} {\bibfnamefont {K.}~\bibnamefont {Melnikov}},
  \ and\ \bibinfo {author} {\bibfnamefont {F.}~\bibnamefont {Petriello}},\
  }\href {\doibase 10.1103/PhysRevD.85.034025} {\bibfield  {journal} {\bibinfo
  {journal} {Phys.Rev.}\ }\textbf {\bibinfo {volume} {D85}},\ \bibinfo {pages}
  {034025} (\bibinfo {year} {2012})},\ \Eprint {http://arxiv.org/abs/1111.7041}
  {arXiv:1111.7041 [hep-ph]} \BibitemShut {NoStop}%
\bibitem [{\citenamefont {Garland}\ \emph
  {et~al.}(2002{\natexlab{a}})\citenamefont {Garland}, \citenamefont
  {Gehrmann}, \citenamefont {Glover}, \citenamefont {Koukoutsakis},\ and\
  \citenamefont {Remiddi}}]{Garland:2001tf}%
  \BibitemOpen
  \bibfield  {author} {\bibinfo {author} {\bibfnamefont {L.}~\bibnamefont
  {Garland}}, \bibinfo {author} {\bibfnamefont {T.}~\bibnamefont {Gehrmann}},
  \bibinfo {author} {\bibfnamefont {E.~N.}\ \bibnamefont {Glover}}, \bibinfo
  {author} {\bibfnamefont {A.}~\bibnamefont {Koukoutsakis}}, \ and\ \bibinfo
  {author} {\bibfnamefont {E.}~\bibnamefont {Remiddi}},\ }\href {\doibase
  10.1016/S0550-3213(02)00057-3} {\bibfield  {journal} {\bibinfo  {journal}
  {Nucl.Phys.}\ }\textbf {\bibinfo {volume} {B627}},\ \bibinfo {pages} {107}
  (\bibinfo {year} {2002}{\natexlab{a}})},\ \Eprint
  {http://arxiv.org/abs/hep-ph/0112081} {arXiv:hep-ph/0112081 [hep-ph]}
  \BibitemShut {NoStop}%
\bibitem [{\citenamefont {Garland}\ \emph
  {et~al.}(2002{\natexlab{b}})\citenamefont {Garland}, \citenamefont
  {Gehrmann}, \citenamefont {Glover}, \citenamefont {Koukoutsakis},\ and\
  \citenamefont {Remiddi}}]{Garland:2002ak}%
  \BibitemOpen
  \bibfield  {author} {\bibinfo {author} {\bibfnamefont {L.}~\bibnamefont
  {Garland}}, \bibinfo {author} {\bibfnamefont {T.}~\bibnamefont {Gehrmann}},
  \bibinfo {author} {\bibfnamefont {E.~W.~N.}\ \bibnamefont {Glover}}, \bibinfo
  {author} {\bibfnamefont {A.}~\bibnamefont {Koukoutsakis}}, \ and\ \bibinfo
  {author} {\bibfnamefont {E.}~\bibnamefont {Remiddi}},\ }\href {\doibase
  10.1016/S0550-3213(02)00627-2} {\bibfield  {journal} {\bibinfo  {journal}
  {Nucl.Phys.}\ }\textbf {\bibinfo {volume} {B642}},\ \bibinfo {pages} {227}
  (\bibinfo {year} {2002}{\natexlab{b}})},\ \Eprint
  {http://arxiv.org/abs/hep-ph/0206067} {arXiv:hep-ph/0206067 [hep-ph]}
  \BibitemShut {NoStop}%
\bibitem [{\citenamefont {Gehrmann}\ \emph {et~al.}(2012)\citenamefont
  {Gehrmann}, \citenamefont {Jaquier}, \citenamefont {Glover},\ and\
  \citenamefont {Koukoutsakis}}]{Gehrmann:2011aa}%
  \BibitemOpen
  \bibfield  {author} {\bibinfo {author} {\bibfnamefont {T.}~\bibnamefont
  {Gehrmann}}, \bibinfo {author} {\bibfnamefont {M.}~\bibnamefont {Jaquier}},
  \bibinfo {author} {\bibfnamefont {E.}~\bibnamefont {Glover}}, \ and\ \bibinfo
  {author} {\bibfnamefont {A.}~\bibnamefont {Koukoutsakis}},\ }\href {\doibase
  10.1007/JHEP02(2012)056} {\bibfield  {journal} {\bibinfo  {journal} {JHEP}\
  }\textbf {\bibinfo {volume} {1202}},\ \bibinfo {pages} {056} (\bibinfo {year}
  {2012})},\ \Eprint {http://arxiv.org/abs/1112.3554} {arXiv:1112.3554
  [hep-ph]} \BibitemShut {NoStop}%
\bibitem [{\citenamefont {Bern}\ \emph {et~al.}(2005)\citenamefont {Bern},
  \citenamefont {Dixon},\ and\ \citenamefont {Smirnov}}]{Bern:2005iz}%
  \BibitemOpen
  \bibfield  {author} {\bibinfo {author} {\bibfnamefont {Z.}~\bibnamefont
  {Bern}}, \bibinfo {author} {\bibfnamefont {L.~J.}\ \bibnamefont {Dixon}}, \
  and\ \bibinfo {author} {\bibfnamefont {V.~A.}\ \bibnamefont {Smirnov}},\
  }\href {\doibase 10.1103/PhysRevD.72.085001} {\bibfield  {journal} {\bibinfo
  {journal} {Phys.Rev.}\ }\textbf {\bibinfo {volume} {D72}},\ \bibinfo {pages}
  {085001} (\bibinfo {year} {2005})},\ \Eprint
  {http://arxiv.org/abs/hep-th/0505205} {arXiv:hep-th/0505205 [hep-th]}
  \BibitemShut {NoStop}%
\bibitem [{\citenamefont {Spradlin}\ \emph {et~al.}(2008)\citenamefont
  {Spradlin}, \citenamefont {Volovich},\ and\ \citenamefont
  {Wen}}]{Spradlin:2008uu}%
  \BibitemOpen
  \bibfield  {author} {\bibinfo {author} {\bibfnamefont {M.}~\bibnamefont
  {Spradlin}}, \bibinfo {author} {\bibfnamefont {A.}~\bibnamefont {Volovich}},
  \ and\ \bibinfo {author} {\bibfnamefont {C.}~\bibnamefont {Wen}},\ }\href
  {\doibase 10.1103/PhysRevD.78.085025} {\bibfield  {journal} {\bibinfo
  {journal} {Phys.Rev.}\ }\textbf {\bibinfo {volume} {D78}},\ \bibinfo {pages}
  {085025} (\bibinfo {year} {2008})},\ \Eprint {http://arxiv.org/abs/0808.1054}
  {arXiv:0808.1054 [hep-th]} \BibitemShut {NoStop}%
\bibitem [{\citenamefont {Gatheral}(1983)}]{Gatheral:1983cz}%
  \BibitemOpen
  \bibfield  {author} {\bibinfo {author} {\bibfnamefont {J.}~\bibnamefont
  {Gatheral}},\ }\href {\doibase 10.1016/0370-2693(83)90112-0} {\bibfield
  {journal} {\bibinfo  {journal} {Phys.Lett.}\ }\textbf {\bibinfo {volume}
  {B133}},\ \bibinfo {pages} {90} (\bibinfo {year} {1983})}\BibitemShut
  {NoStop}%
\bibitem [{\citenamefont {Frenkel}\ and\ \citenamefont
  {Taylor}(1984)}]{Frenkel:1984pz}%
  \BibitemOpen
  \bibfield  {author} {\bibinfo {author} {\bibfnamefont {J.}~\bibnamefont
  {Frenkel}}\ and\ \bibinfo {author} {\bibfnamefont {J.}~\bibnamefont
  {Taylor}},\ }\href {\doibase 10.1016/0550-3213(84)90294-3} {\bibfield
  {journal} {\bibinfo  {journal} {Nucl.Phys.}\ }\textbf {\bibinfo {volume}
  {B246}},\ \bibinfo {pages} {231} (\bibinfo {year} {1984})}\BibitemShut
  {NoStop}%
\bibitem [{\citenamefont {Catani}\ \emph {et~al.}(2001)\citenamefont {Catani},
  \citenamefont {de~Florian},\ and\ \citenamefont {Grazzini}}]{Catani:2001ic}%
  \BibitemOpen
  \bibfield  {author} {\bibinfo {author} {\bibfnamefont {S.}~\bibnamefont
  {Catani}}, \bibinfo {author} {\bibfnamefont {D.}~\bibnamefont {de~Florian}},
  \ and\ \bibinfo {author} {\bibfnamefont {M.}~\bibnamefont {Grazzini}},\
  }\href@noop {} {\bibfield  {journal} {\bibinfo  {journal} {JHEP}\ }\textbf
  {\bibinfo {volume} {0105}},\ \bibinfo {pages} {025} (\bibinfo {year}
  {2001})},\ \Eprint {http://arxiv.org/abs/hep-ph/0102227}
  {arXiv:hep-ph/0102227 [hep-ph]} \BibitemShut {NoStop}%
\bibitem [{\citenamefont {Harlander}\ and\ \citenamefont
  {Kilgore}(2001)}]{Harlander:2001is}%
  \BibitemOpen
  \bibfield  {author} {\bibinfo {author} {\bibfnamefont {R.~V.}\ \bibnamefont
  {Harlander}}\ and\ \bibinfo {author} {\bibfnamefont {W.~B.}\ \bibnamefont
  {Kilgore}},\ }\href {\doibase 10.1103/PhysRevD.64.013015} {\bibfield
  {journal} {\bibinfo  {journal} {Phys.Rev.}\ }\textbf {\bibinfo {volume}
  {D64}},\ \bibinfo {pages} {013015} (\bibinfo {year} {2001})},\ \Eprint
  {http://arxiv.org/abs/hep-ph/0102241} {arXiv:hep-ph/0102241 [hep-ph]}
  \BibitemShut {NoStop}%
\bibitem [{\citenamefont {de~Florian}\ and\ \citenamefont
  {Mazzitelli}(2012)}]{deFlorian:2012za}%
  \BibitemOpen
  \bibfield  {author} {\bibinfo {author} {\bibfnamefont {D.}~\bibnamefont
  {de~Florian}}\ and\ \bibinfo {author} {\bibfnamefont {J.}~\bibnamefont
  {Mazzitelli}},\ }\href {\doibase 10.1007/JHEP12(2012)08} {\bibfield
  {journal} {\bibinfo  {journal} {JHEP}\ }\textbf {\bibinfo {volume} {1212}},\
  \bibinfo {pages} {088} (\bibinfo {year} {2012})},\ \Eprint
  {http://arxiv.org/abs/1209.0673} {arXiv:1209.0673 [hep-ph]} \BibitemShut
  {NoStop}%
\bibitem [{\citenamefont {Bauer}\ \emph {et~al.}(2000)\citenamefont {Bauer},
  \citenamefont {Fleming},\ and\ \citenamefont {Luke}}]{Bauer:2000ew}%
  \BibitemOpen
  \bibfield  {author} {\bibinfo {author} {\bibfnamefont {C.~W.}\ \bibnamefont
  {Bauer}}, \bibinfo {author} {\bibfnamefont {S.}~\bibnamefont {Fleming}}, \
  and\ \bibinfo {author} {\bibfnamefont {M.~E.}\ \bibnamefont {Luke}},\ }\href
  {\doibase 10.1103/PhysRevD.63.014006} {\bibfield  {journal} {\bibinfo
  {journal} {Phys.Rev.}\ }\textbf {\bibinfo {volume} {D63}},\ \bibinfo {pages}
  {014006} (\bibinfo {year} {2000})},\ \Eprint
  {http://arxiv.org/abs/hep-ph/0005275} {arXiv:hep-ph/0005275 [hep-ph]}
  \BibitemShut {NoStop}%
\bibitem [{\citenamefont {Bauer}\ \emph {et~al.}(2001)\citenamefont {Bauer},
  \citenamefont {Fleming}, \citenamefont {Pirjol},\ and\ \citenamefont
  {Stewart}}]{Bauer:2000yr}%
  \BibitemOpen
  \bibfield  {author} {\bibinfo {author} {\bibfnamefont {C.~W.}\ \bibnamefont
  {Bauer}}, \bibinfo {author} {\bibfnamefont {S.}~\bibnamefont {Fleming}},
  \bibinfo {author} {\bibfnamefont {D.}~\bibnamefont {Pirjol}}, \ and\ \bibinfo
  {author} {\bibfnamefont {I.~W.}\ \bibnamefont {Stewart}},\ }\href {\doibase
  10.1103/PhysRevD.63.114020} {\bibfield  {journal} {\bibinfo  {journal}
  {Phys.Rev.}\ }\textbf {\bibinfo {volume} {D63}},\ \bibinfo {pages} {114020}
  (\bibinfo {year} {2001})},\ \Eprint {http://arxiv.org/abs/hep-ph/0011336}
  {arXiv:hep-ph/0011336 [hep-ph]} \BibitemShut {NoStop}%
\bibitem [{\citenamefont {Bauer}\ \emph {et~al.}(2002)\citenamefont {Bauer},
  \citenamefont {Pirjol},\ and\ \citenamefont {Stewart}}]{Bauer:2001yt}%
  \BibitemOpen
  \bibfield  {author} {\bibinfo {author} {\bibfnamefont {C.~W.}\ \bibnamefont
  {Bauer}}, \bibinfo {author} {\bibfnamefont {D.}~\bibnamefont {Pirjol}}, \
  and\ \bibinfo {author} {\bibfnamefont {I.~W.}\ \bibnamefont {Stewart}},\
  }\href {\doibase 10.1103/PhysRevD.65.054022} {\bibfield  {journal} {\bibinfo
  {journal} {Phys.Rev.}\ }\textbf {\bibinfo {volume} {D65}},\ \bibinfo {pages}
  {054022} (\bibinfo {year} {2002})},\ \Eprint
  {http://arxiv.org/abs/hep-ph/0109045} {arXiv:hep-ph/0109045 [hep-ph]}
  \BibitemShut {NoStop}%
\bibitem [{\citenamefont {Beneke}\ \emph {et~al.}(2002)\citenamefont {Beneke},
  \citenamefont {Chapovsky}, \citenamefont {Diehl},\ and\ \citenamefont
  {Feldmann}}]{Beneke:2002ph}%
  \BibitemOpen
  \bibfield  {author} {\bibinfo {author} {\bibfnamefont {M.}~\bibnamefont
  {Beneke}}, \bibinfo {author} {\bibfnamefont {A.}~\bibnamefont {Chapovsky}},
  \bibinfo {author} {\bibfnamefont {M.}~\bibnamefont {Diehl}}, \ and\ \bibinfo
  {author} {\bibfnamefont {T.}~\bibnamefont {Feldmann}},\ }\href {\doibase
  10.1016/S0550-3213(02)00687-9} {\bibfield  {journal} {\bibinfo  {journal}
  {Nucl.Phys.}\ }\textbf {\bibinfo {volume} {B643}},\ \bibinfo {pages} {431}
  (\bibinfo {year} {2002})},\ \Eprint {http://arxiv.org/abs/hep-ph/0206152}
  {arXiv:hep-ph/0206152 [hep-ph]} \BibitemShut {NoStop}%
\bibitem [{\citenamefont {Nogueira}(1993)}]{Nogueira:1991ex}%
  \BibitemOpen
  \bibfield  {author} {\bibinfo {author} {\bibfnamefont {P.}~\bibnamefont
  {Nogueira}},\ }\href {\doibase 10.1006/jcph.1993.1074} {\bibfield  {journal}
  {\bibinfo  {journal} {J.Comput.Phys.}\ }\textbf {\bibinfo {volume} {105}},\
  \bibinfo {pages} {279} (\bibinfo {year} {1993})}\BibitemShut {NoStop}%
\bibitem [{\citenamefont {Chetyrkin}\ and\ \citenamefont
  {Tkachov}(1981)}]{Chetyrkin:1981qh}%
  \BibitemOpen
  \bibfield  {author} {\bibinfo {author} {\bibfnamefont {K.}~\bibnamefont
  {Chetyrkin}}\ and\ \bibinfo {author} {\bibfnamefont {F.}~\bibnamefont
  {Tkachov}},\ }\href {\doibase 10.1016/0550-3213(81)90199-1} {\bibfield
  {journal} {\bibinfo  {journal} {Nucl.Phys.}\ }\textbf {\bibinfo {volume}
  {B192}},\ \bibinfo {pages} {159} (\bibinfo {year} {1981})}\BibitemShut
  {NoStop}%
\bibitem [{\citenamefont {Tkachov}(1981)}]{Tkachov:1981wb}%
  \BibitemOpen
  \bibfield  {author} {\bibinfo {author} {\bibfnamefont {F.}~\bibnamefont
  {Tkachov}},\ }\href {\doibase 10.1016/0370-2693(81)90288-4} {\bibfield
  {journal} {\bibinfo  {journal} {Phys.Lett.}\ }\textbf {\bibinfo {volume}
  {B100}},\ \bibinfo {pages} {65} (\bibinfo {year} {1981})}\BibitemShut
  {NoStop}%
\bibitem [{\citenamefont {Smirnov}(2008)}]{Smirnov:2008iw}%
  \BibitemOpen
  \bibfield  {author} {\bibinfo {author} {\bibfnamefont {A.}~\bibnamefont
  {Smirnov}},\ }\href {\doibase 10.1088/1126-6708/2008/10/107} {\bibfield
  {journal} {\bibinfo  {journal} {JHEP}\ }\textbf {\bibinfo {volume} {0810}},\
  \bibinfo {pages} {107} (\bibinfo {year} {2008})},\ \Eprint
  {http://arxiv.org/abs/0807.3243} {arXiv:0807.3243 [hep-ph]} \BibitemShut
  {NoStop}%
\bibitem [{\citenamefont {Laporta}(2000)}]{Laporta:2001dd}%
  \BibitemOpen
  \bibfield  {author} {\bibinfo {author} {\bibfnamefont {S.}~\bibnamefont
  {Laporta}},\ }\href {\doibase 10.1016/S0217-751X(00)00215-7} {\bibfield
  {journal} {\bibinfo  {journal} {Int.J.Mod.Phys.}\ }\textbf {\bibinfo {volume}
  {A15}},\ \bibinfo {pages} {5087} (\bibinfo {year} {2000})},\ \Eprint
  {http://arxiv.org/abs/hep-ph/0102033} {arXiv:hep-ph/0102033 [hep-ph]}
  \BibitemShut {NoStop}%
\bibitem [{\citenamefont {Lee}(2012)}]{Lee:2012cn}%
  \BibitemOpen
  \bibfield  {author} {\bibinfo {author} {\bibfnamefont {R.}~\bibnamefont
  {Lee}},\ }\href@noop {} {\  (\bibinfo {year} {2012})},\ \Eprint
  {http://arxiv.org/abs/1212.2685} {arXiv:1212.2685 [hep-ph]} \BibitemShut
  {NoStop}%
\bibitem [{\citenamefont {Bern}\ and\ \citenamefont
  {Kosower}(1992)}]{Bern:1991aq}%
  \BibitemOpen
  \bibfield  {author} {\bibinfo {author} {\bibfnamefont {Z.}~\bibnamefont
  {Bern}}\ and\ \bibinfo {author} {\bibfnamefont {D.~A.}\ \bibnamefont
  {Kosower}},\ }\href {\doibase 10.1016/0550-3213(92)90134-W} {\bibfield
  {journal} {\bibinfo  {journal} {Nucl.Phys.}\ }\textbf {\bibinfo {volume}
  {B379}},\ \bibinfo {pages} {451} (\bibinfo {year} {1992})}\BibitemShut
  {NoStop}%
\bibitem [{\citenamefont {Bern}\ \emph {et~al.}(2002)\citenamefont {Bern},
  \citenamefont {De~Freitas}, \citenamefont {Dixon},\ and\ \citenamefont
  {Wong}}]{Bern:2002zk}%
  \BibitemOpen
  \bibfield  {author} {\bibinfo {author} {\bibfnamefont {Z.}~\bibnamefont
  {Bern}}, \bibinfo {author} {\bibfnamefont {A.}~\bibnamefont {De~Freitas}},
  \bibinfo {author} {\bibfnamefont {L.~J.}\ \bibnamefont {Dixon}}, \ and\
  \bibinfo {author} {\bibfnamefont {H.}~\bibnamefont {Wong}},\ }\href {\doibase
  10.1103/PhysRevD.66.085002} {\bibfield  {journal} {\bibinfo  {journal}
  {Phys.Rev.}\ }\textbf {\bibinfo {volume} {D66}},\ \bibinfo {pages} {085002}
  (\bibinfo {year} {2002})},\ \Eprint {http://arxiv.org/abs/hep-ph/0202271}
  {arXiv:hep-ph/0202271 [hep-ph]} \BibitemShut {NoStop}%
\bibitem [{\citenamefont {Gehrmann}\ and\ \citenamefont
  {Remiddi}(2001)}]{Gehrmann:2000zt}%
  \BibitemOpen
  \bibfield  {author} {\bibinfo {author} {\bibfnamefont {T.}~\bibnamefont
  {Gehrmann}}\ and\ \bibinfo {author} {\bibfnamefont {E.}~\bibnamefont
  {Remiddi}},\ }\href {\doibase 10.1016/S0550-3213(01)00057-8} {\bibfield
  {journal} {\bibinfo  {journal} {Nucl.Phys.}\ }\textbf {\bibinfo {volume}
  {B601}},\ \bibinfo {pages} {248} (\bibinfo {year} {2001})},\ \Eprint
  {http://arxiv.org/abs/hep-ph/0008287} {arXiv:hep-ph/0008287 [hep-ph]}
  \BibitemShut {NoStop}%
\bibitem [{\citenamefont {Czakon}(2006)}]{Czakon:2005rk}%
  \BibitemOpen
  \bibfield  {author} {\bibinfo {author} {\bibfnamefont {M.}~\bibnamefont
  {Czakon}},\ }\href {\doibase 10.1016/j.cpc.2006.07.002} {\bibfield  {journal}
  {\bibinfo  {journal} {Comput.Phys.Commun.}\ }\textbf {\bibinfo {volume}
  {175}},\ \bibinfo {pages} {559} (\bibinfo {year} {2006})},\ \Eprint
  {http://arxiv.org/abs/hep-ph/0511200} {arXiv:hep-ph/0511200 [hep-ph]}
  \BibitemShut {NoStop}%
\bibitem [{\citenamefont {Kotikov}\ \emph {et~al.}(2004)\citenamefont
  {Kotikov}, \citenamefont {Lipatov}, \citenamefont {Onishchenko},\ and\
  \citenamefont {Velizhanin}}]{Kotikov:2004er}%
  \BibitemOpen
  \bibfield  {author} {\bibinfo {author} {\bibfnamefont {A.}~\bibnamefont
  {Kotikov}}, \bibinfo {author} {\bibfnamefont {L.}~\bibnamefont {Lipatov}},
  \bibinfo {author} {\bibfnamefont {A.}~\bibnamefont {Onishchenko}}, \ and\
  \bibinfo {author} {\bibfnamefont {V.}~\bibnamefont {Velizhanin}},\ }\href
  {\doibase 10.1016/j.physletb.2004.05.078} {\bibfield  {journal} {\bibinfo
  {journal} {Phys.Lett.}\ }\textbf {\bibinfo {volume} {B595}},\ \bibinfo
  {pages} {521} (\bibinfo {year} {2004})},\ \Eprint
  {http://arxiv.org/abs/hep-th/0404092} {arXiv:hep-th/0404092 [hep-th]}
  \BibitemShut {NoStop}%
\bibitem [{\citenamefont {Anastasiou}\ \emph {et~al.}(2003)\citenamefont
  {Anastasiou}, \citenamefont {Bern}, \citenamefont {Dixon},\ and\
  \citenamefont {Kosower}}]{Anastasiou:2003kj}%
  \BibitemOpen
  \bibfield  {author} {\bibinfo {author} {\bibfnamefont {C.}~\bibnamefont
  {Anastasiou}}, \bibinfo {author} {\bibfnamefont {Z.}~\bibnamefont {Bern}},
  \bibinfo {author} {\bibfnamefont {L.~J.}\ \bibnamefont {Dixon}}, \ and\
  \bibinfo {author} {\bibfnamefont {D.}~\bibnamefont {Kosower}},\ }\href
  {\doibase 10.1103/PhysRevLett.91.251602} {\bibfield  {journal} {\bibinfo
  {journal} {Phys.Rev.Lett.}\ }\textbf {\bibinfo {volume} {91}},\ \bibinfo
  {pages} {251602} (\bibinfo {year} {2003})},\ \Eprint
  {http://arxiv.org/abs/hep-th/0309040} {arXiv:hep-th/0309040 [hep-th]}
  \BibitemShut {NoStop}%
\bibitem [{\citenamefont {Bern}\ \emph {et~al.}(2007)\citenamefont {Bern},
  \citenamefont {Czakon}, \citenamefont {Dixon}, \citenamefont {Kosower},\ and\
  \citenamefont {Smirnov}}]{Bern:2006ew}%
  \BibitemOpen
  \bibfield  {author} {\bibinfo {author} {\bibfnamefont {Z.}~\bibnamefont
  {Bern}}, \bibinfo {author} {\bibfnamefont {M.}~\bibnamefont {Czakon}},
  \bibinfo {author} {\bibfnamefont {L.~J.}\ \bibnamefont {Dixon}}, \bibinfo
  {author} {\bibfnamefont {D.~A.}\ \bibnamefont {Kosower}}, \ and\ \bibinfo
  {author} {\bibfnamefont {V.~A.}\ \bibnamefont {Smirnov}},\ }\href {\doibase
  10.1103/PhysRevD.75.085010} {\bibfield  {journal} {\bibinfo  {journal}
  {Phys.Rev.}\ }\textbf {\bibinfo {volume} {D75}},\ \bibinfo {pages} {085010}
  (\bibinfo {year} {2007})},\ \Eprint {http://arxiv.org/abs/hep-th/0610248}
  {arXiv:hep-th/0610248 [hep-th]} \BibitemShut {NoStop}%
\bibitem [{\citenamefont {Pak}\ \emph {et~al.}(2011)\citenamefont {Pak},
  \citenamefont {Rogal},\ and\ \citenamefont {Steinhauser}}]{Pak:2011hs}%
  \BibitemOpen
  \bibfield  {author} {\bibinfo {author} {\bibfnamefont {A.}~\bibnamefont
  {Pak}}, \bibinfo {author} {\bibfnamefont {M.}~\bibnamefont {Rogal}}, \ and\
  \bibinfo {author} {\bibfnamefont {M.}~\bibnamefont {Steinhauser}},\ }\href
  {\doibase 10.1007/JHEP09(2011)088} {\bibfield  {journal} {\bibinfo  {journal}
  {JHEP}\ }\textbf {\bibinfo {volume} {1109}},\ \bibinfo {pages} {088}
  (\bibinfo {year} {2011})},\ \Eprint {http://arxiv.org/abs/1107.3391}
  {arXiv:1107.3391 [hep-ph]} \BibitemShut {NoStop}%
\bibitem [{\citenamefont {Anastasiou}\ \emph {et~al.}(2012)\citenamefont
  {Anastasiou}, \citenamefont {Buehler}, \citenamefont {Duhr},\ and\
  \citenamefont {Herzog}}]{Anastasiou:2012kq}%
  \BibitemOpen
  \bibfield  {author} {\bibinfo {author} {\bibfnamefont {C.}~\bibnamefont
  {Anastasiou}}, \bibinfo {author} {\bibfnamefont {S.}~\bibnamefont {Buehler}},
  \bibinfo {author} {\bibfnamefont {C.}~\bibnamefont {Duhr}}, \ and\ \bibinfo
  {author} {\bibfnamefont {F.}~\bibnamefont {Herzog}},\ }\href {\doibase
  10.1007/JHEP11(2012)062} {\bibfield  {journal} {\bibinfo  {journal} {JHEP}\
  }\textbf {\bibinfo {volume} {1211}},\ \bibinfo {pages} {062} (\bibinfo {year}
  {2012})},\ \Eprint {http://arxiv.org/abs/1208.3130} {arXiv:1208.3130
  [hep-ph]} \BibitemShut {NoStop}%
\bibitem [{\citenamefont {Höschele}\ \emph {et~al.}(2013)\citenamefont
  {Höschele}, \citenamefont {Hoff}, \citenamefont {Pak}, \citenamefont
  {Steinhauser},\ and\ \citenamefont {Ueda}}]{Hoschele:2012xc}%
  \BibitemOpen
  \bibfield  {author} {\bibinfo {author} {\bibfnamefont {M.}~\bibnamefont
  {Höschele}}, \bibinfo {author} {\bibfnamefont {J.}~\bibnamefont {Hoff}},
  \bibinfo {author} {\bibfnamefont {A.}~\bibnamefont {Pak}}, \bibinfo {author}
  {\bibfnamefont {M.}~\bibnamefont {Steinhauser}}, \ and\ \bibinfo {author}
  {\bibfnamefont {T.}~\bibnamefont {Ueda}},\ }\href {\doibase
  10.1016/j.physletb.2013.03.003} {\bibfield  {journal} {\bibinfo  {journal}
  {Phys.Lett.}\ }\textbf {\bibinfo {volume} {B721}},\ \bibinfo {pages} {244}
  (\bibinfo {year} {2013})},\ \Eprint {http://arxiv.org/abs/1211.6559}
  {arXiv:1211.6559 [hep-ph]} \BibitemShut {NoStop}%
\bibitem [{\citenamefont {Anastasiou}\ \emph {et~al.}(2013)\citenamefont
  {Anastasiou}, \citenamefont {Duhr}, \citenamefont {Dulat},\ and\
  \citenamefont {Mistlberger}}]{Anastasiou:2013srw}%
  \BibitemOpen
  \bibfield  {author} {\bibinfo {author} {\bibfnamefont {C.}~\bibnamefont
  {Anastasiou}}, \bibinfo {author} {\bibfnamefont {C.}~\bibnamefont {Duhr}},
  \bibinfo {author} {\bibfnamefont {F.}~\bibnamefont {Dulat}}, \ and\ \bibinfo
  {author} {\bibfnamefont {B.}~\bibnamefont {Mistlberger}},\ }\href {\doibase
  10.1007/JHEP07(2013)003} {\bibfield  {journal} {\bibinfo  {journal} {JHEP}\
  }\textbf {\bibinfo {volume} {1307}},\ \bibinfo {pages} {003} (\bibinfo {year}
  {2013})},\ \Eprint {http://arxiv.org/abs/1302.4379} {arXiv:1302.4379
  [hep-ph]} \BibitemShut {NoStop}%
\bibitem [{\citenamefont {Ball}\ \emph {et~al.}(2013)\citenamefont {Ball},
  \citenamefont {Bonvini}, \citenamefont {Forte}, \citenamefont {Marzani},\
  and\ \citenamefont {Ridolfi}}]{Ball:2013bra}%
  \BibitemOpen
  \bibfield  {author} {\bibinfo {author} {\bibfnamefont {R.~D.}\ \bibnamefont
  {Ball}}, \bibinfo {author} {\bibfnamefont {M.}~\bibnamefont {Bonvini}},
  \bibinfo {author} {\bibfnamefont {S.}~\bibnamefont {Forte}}, \bibinfo
  {author} {\bibfnamefont {S.}~\bibnamefont {Marzani}}, \ and\ \bibinfo
  {author} {\bibfnamefont {G.}~\bibnamefont {Ridolfi}},\ }\href {\doibase
  10.1016/j.nuclphysb.2013.06.012} {\bibfield  {journal} {\bibinfo  {journal}
  {Nucl.Phys.}\ }\textbf {\bibinfo {volume} {B874}},\ \bibinfo {pages} {746}
  (\bibinfo {year} {2013})},\ \Eprint {http://arxiv.org/abs/1303.3590}
  {arXiv:1303.3590 [hep-ph]} \BibitemShut {NoStop}%
\bibitem [{\citenamefont {Buehler}\ and\ \citenamefont
  {Lazopoulos}(2013)}]{Buehler:2013fha}%
  \BibitemOpen
  \bibfield  {author} {\bibinfo {author} {\bibfnamefont {S.}~\bibnamefont
  {Buehler}}\ and\ \bibinfo {author} {\bibfnamefont {A.}~\bibnamefont
  {Lazopoulos}},\ }\href@noop {} {\  (\bibinfo {year} {2013})},\ \Eprint
  {http://arxiv.org/abs/1306.2223} {arXiv:1306.2223 [hep-ph]} \BibitemShut
  {NoStop}%
\bibitem [{\citenamefont {Gardi}\ \emph {et~al.}(2013)\citenamefont {Gardi},
  \citenamefont {Smillie},\ and\ \citenamefont {White}}]{Gardi:2013ita}%
  \BibitemOpen
  \bibfield  {author} {\bibinfo {author} {\bibfnamefont {E.}~\bibnamefont
  {Gardi}}, \bibinfo {author} {\bibfnamefont {J.~M.}\ \bibnamefont {Smillie}},
  \ and\ \bibinfo {author} {\bibfnamefont {C.~D.}\ \bibnamefont {White}},\
  }\href@noop {} {\  (\bibinfo {year} {2013})},\ \Eprint
  {http://arxiv.org/abs/1304.7040} {arXiv:1304.7040 [hep-ph]} \BibitemShut
  {NoStop}%
\bibitem [{\citenamefont {Duhr}\ and\ \citenamefont
  {Gehrmann}(2013)}]{Duhr:2013msa}%
  \BibitemOpen
  \bibfield  {author} {\bibinfo {author} {\bibfnamefont {C.}~\bibnamefont
  {Duhr}}\ and\ \bibinfo {author} {\bibfnamefont {T.}~\bibnamefont
  {Gehrmann}},\ }\href@noop {} {\  (\bibinfo {year} {2013})},\ \Eprint
  {http://arxiv.org/abs/1309.4393} {arXiv:1309.4393 [hep-ph]} \BibitemShut
  {NoStop}%
\bibitem [{\citenamefont {Ferguson}\ and\ \citenamefont
  {Bailey}(1991)}]{pslq1}%
  \BibitemOpen
  \bibfield  {author} {\bibinfo {author} {\bibfnamefont {H.~R.~P.}\
  \bibnamefont {Ferguson}}\ and\ \bibinfo {author} {\bibfnamefont {D.~H.}\
  \bibnamefont {Bailey}},\ }\href@noop {} {\bibfield  {journal} {\bibinfo
  {journal} {RNR Technical Report RNR-91-032}\ } (\bibinfo {year}
  {1991})}\BibitemShut {NoStop}%
\bibitem [{\citenamefont {Ferguson}\ \emph {et~al.}(1999)\citenamefont
  {Ferguson}, \citenamefont {Bailey},\ and\ \citenamefont {Arno}}]{pslq2}%
  \BibitemOpen
  \bibfield  {author} {\bibinfo {author} {\bibfnamefont {H.~R.~P.}\
  \bibnamefont {Ferguson}}, \bibinfo {author} {\bibfnamefont {D.~H.}\
  \bibnamefont {Bailey}}, \ and\ \bibinfo {author} {\bibfnamefont
  {S.}~\bibnamefont {Arno}},\ }\href@noop {} {\bibfield  {journal} {\bibinfo
  {journal} {Math. Comput.}\ }\textbf {\bibinfo {volume} {68}},\ \bibinfo
  {pages} {351} (\bibinfo {year} {1999})}\BibitemShut {NoStop}%
\bibitem [{\citenamefont {Vermaseren}(1999)}]{Vermaseren:1998uu}%
  \BibitemOpen
  \bibfield  {author} {\bibinfo {author} {\bibfnamefont {J.}~\bibnamefont
  {Vermaseren}},\ }\href {\doibase 10.1142/S0217751X99001032} {\bibfield
  {journal} {\bibinfo  {journal} {Int.J.Mod.Phys.}\ }\textbf {\bibinfo {volume}
  {A14}},\ \bibinfo {pages} {2037} (\bibinfo {year} {1999})},\ \Eprint
  {http://arxiv.org/abs/hep-ph/9806280} {arXiv:hep-ph/9806280 [hep-ph]}
  \BibitemShut {NoStop}%
\bibitem [{\citenamefont {Moch}\ and\ \citenamefont
  {Uwer}(2006)}]{Moch:2005uc}%
  \BibitemOpen
  \bibfield  {author} {\bibinfo {author} {\bibfnamefont {S.}~\bibnamefont
  {Moch}}\ and\ \bibinfo {author} {\bibfnamefont {P.}~\bibnamefont {Uwer}},\
  }\href {\doibase 10.1016/j.cpc.2005.12.014} {\bibfield  {journal} {\bibinfo
  {journal} {Comput.Phys.Commun.}\ }\textbf {\bibinfo {volume} {174}},\
  \bibinfo {pages} {759} (\bibinfo {year} {2006})},\ \Eprint
  {http://arxiv.org/abs/math-ph/0508008} {arXiv:math-ph/0508008 [math-ph]}
  \BibitemShut {NoStop}%
\bibitem [{\citenamefont {Smirnov}\ \emph {et~al.}(2011)\citenamefont
  {Smirnov}, \citenamefont {Smirnov},\ and\ \citenamefont
  {Tentyukov}}]{Smirnov:2009pb}%
  \BibitemOpen
  \bibfield  {author} {\bibinfo {author} {\bibfnamefont {A.}~\bibnamefont
  {Smirnov}}, \bibinfo {author} {\bibfnamefont {V.}~\bibnamefont {Smirnov}}, \
  and\ \bibinfo {author} {\bibfnamefont {M.}~\bibnamefont {Tentyukov}},\ }\href
  {\doibase 10.1016/j.cpc.2010.11.025} {\bibfield  {journal} {\bibinfo
  {journal} {Comput.Phys.Commun.}\ }\textbf {\bibinfo {volume} {182}},\
  \bibinfo {pages} {790} (\bibinfo {year} {2011})},\ \Eprint
  {http://arxiv.org/abs/0912.0158} {arXiv:0912.0158 [hep-ph]} \BibitemShut
  {NoStop}%
\bibitem [{\citenamefont {Binoth}\ and\ \citenamefont
  {Heinrich}(2000)}]{Binoth:2000ps}%
  \BibitemOpen
  \bibfield  {author} {\bibinfo {author} {\bibfnamefont {T.}~\bibnamefont
  {Binoth}}\ and\ \bibinfo {author} {\bibfnamefont {G.}~\bibnamefont
  {Heinrich}},\ }\href {\doibase 10.1016/S0550-3213(00)00429-6} {\bibfield
  {journal} {\bibinfo  {journal} {Nucl.Phys.}\ }\textbf {\bibinfo {volume}
  {B585}},\ \bibinfo {pages} {741} (\bibinfo {year} {2000})},\ \Eprint
  {http://arxiv.org/abs/hep-ph/0004013} {arXiv:hep-ph/0004013 [hep-ph]}
  \BibitemShut {NoStop}%
\end{thebibliography}%

\end{document}